\documentclass[pdflatex,sn-basic]{sn-jnl}% Basic Springer Nature Reference Style/Chemistry Reference Style

\usepackage{graphicx}%
\usepackage{multirow}%
\usepackage{amsmath,amssymb,amsfonts}%
\usepackage{amsthm}%
\usepackage{mathrsfs}%
\usepackage[title]{appendix}%
\usepackage{xcolor}%
\usepackage{textcomp}%
\usepackage{manyfoot}%
\usepackage{booktabs}%
\usepackage{algorithm}%
\usepackage{algorithmicx}%
\usepackage{algpseudocode}%
\usepackage{listings}%
\usepackage{float}
\usepackage{arydshln}
\usepackage{lscape}
\usepackage{nicefrac}
\begin{document}

\title[]{Nonparametric directional variogram estimation in the presence of outlier blocks}


\author*[1]{\fnm{Jana} \sur{Gierse}}\email{gierse@statistik.tu-dortmund.de}

\author[1]{\fnm{Roland} \sur{Fried}}\email{fried@statistik.tu-dortmund.de}

\affil*[1]{\orgdiv{Department of Statistics}, \orgname{TU Dortmund University}, \orgaddress{\street{Vogelpothsweg 87}, \city{Dortmund}, \postcode{44227}, \state{North Rhine-Westphalia}, \country{Germany}}}


\abstract{This paper proposes robust estimators of the variogram, a statistical tool that is commonly used in geostatistics to capture the spatial dependence structure of data. The new estimators are based on the highly robust minimum covariance determinant estimator and estimate the directional variogram for several lags jointly. Simulations and breakdown considerations confirm the good robustness properties of the new estimators. While Genton's estimator based on the robust estimation of the variance of pairwise sums and differences performs well in case of isolated outliers, the new estimators based on robust estimation of multivariate variance and covariance matrices perform superior to the established alternatives in the presence  of outlier blocks in the data.  The methods are illustrated by an application to satellite data, where outlier blocks may occur because of e.g. clouds.}

\keywords{Breakdown point, MCD estimator, Robustness, Spatial data,  Temporary changes}

\pacs[MSC Classification]{62H11, 86A32}

\newgeometry{bottom = 3cm, left = 2cm, right = 2cm, top = 3cm}
\maketitle

%\bmhead{Statements and Declarations}
%The authors have no interest to declare.

\section{Introduction}\label{ch:intro}

Investigating spatial data, a basic objective is the determination of the spatial dependency between the data. The variogram is a popular measure of spatial dependency and its  estimation is therefore often one of the first steps in spatial data analysis. The estimated variogram can be used for example for kriging, that is, for prediction at unobserved locations.

Assuming $\{Z(\boldsymbol{s}), \boldsymbol{s}\in I\}$ with an index set $I \subset \mathbb{R}^d$ to be an intrinsic stationary random field, its variogram for a given lag $\boldsymbol{h}$ is 
	\begin{align*}
		2\gamma(\boldsymbol{h}) = \text{Var}\left(Z(\boldsymbol{s}) - Z(\boldsymbol{s}+\boldsymbol{h})\right),  \mbox{ for }\boldsymbol{s},\boldsymbol{s}+\boldsymbol{h}\in I.
	\end{align*}
Thereby, intrinsic stationarity means that the increments $Z(\boldsymbol{s})-Z(\boldsymbol{s}+\boldsymbol{h})$ are weakly stationary, with a constant mean and translation-invariant covariance. A common additional assumption is that $\{Z(\boldsymbol{s}), \boldsymbol{s}\in I\}$ is mean stationary.
    
The popular Matheron variogram estimator proposed by \citet{Matheron1962} is defined as
	 \begin{align*}
	2\widehat{\gamma}_M(\boldsymbol{h}) = \frac{1}{|N(\boldsymbol{h})|} \sum_{(\boldsymbol{s}_i, \boldsymbol{s}_j)\in N(\boldsymbol{h})} (Z(\boldsymbol{s}_i) - Z(\boldsymbol{s}_j))^2
	 \end{align*}
with $N(\boldsymbol{h}) = \{(\boldsymbol{s}_i, \boldsymbol{s}_j)\in I^2: \boldsymbol{s}_j - \boldsymbol{s}_i = \boldsymbol{h}\}$ being the set of all pairs of locations with a distance of $\boldsymbol{h}$ and $|N(\boldsymbol{h})|$ being the cardinality of $N(\boldsymbol{h})$. Especially for irregular grids $N(\boldsymbol{h})$ is often extended to contain not only pairs with distance exactly equal to $\boldsymbol{h}$ but also with a distance in a tolerance region $T(\boldsymbol{h})$ around $\boldsymbol{h}$. In case of mean stationarity and without the consideration of tolerance regions $T(\boldsymbol{h})$  this estimator is unbiased and consistent.  However, it is not robust as a single outlier can affect it heavily so that it gets worthless \citep{Genton1998a}.  
  
Because of this deficiency, \citet{Cressie1980} proposed taking the square roots of the differences instead of their squares to reduce the effect of outliers. However, this estimator is not really robust either, since the effect of a single outlier is not bounded \citep{Genton1998a}. More about this estimator and the Matheron estimator e.g. regarding inference can be found in \citet{Bowman2013}.

\citet{Genton1998a} proposes usage of robust scale estimators like the highly robust and yet (under Gaussianity) rather efficient Qn estimator introduced by \citet{Rousseeuw1993} to obtain robust variogram estimators. The resulting Genton variogram estimator is
	\begin{align*}
	2\widehat{\gamma}_G(\boldsymbol{h}) = \left(c \cdot \left(|V_i(\boldsymbol{h}) - V_j(\boldsymbol{h})|: i<j\right)_{(k)}\right)^2
	\end{align*} 
with $k = \begin{pmatrix}\left[\nicefrac{|N(\boldsymbol{h})|}{2}\right] + 1 \\ 2\end{pmatrix}$,  $[a]$ denoting the integer part of $a$, $c$ being a consistency factor, which is approximately equal to $2.22$ in large Gaussian samples, and $(\cdot)_{(k)}$ denotes the $k$-quantile. Furthermore, $V_i(\boldsymbol{h})$ is the $i$-th element of the set $V(\boldsymbol{h}) = \{Z(\boldsymbol{s}_l) - Z(\boldsymbol{s}_m): (\boldsymbol{s}_l, \boldsymbol{s}_m)\in N(\boldsymbol{h})\}$ containing all differences for locations with distance $\boldsymbol{h}$. This estimator, in contrast is robust with a spatial breakdown point of at least 25\% for two-dimensional data. The spatial breakdown point reflects the fact that for spatial data the position of the outliers is important and thus considers the worst spatial configuration for the outliers \citep{Genton1998, Lark2008}. \\
\citet{Lark2000} and \citet{Kerry2007a} compare these estimators in the possible presence of isolated outliers using simulations. Both works verify the robustness of the Genton % and the Dowd 
estimator in such scenarios. Furthermore, they find the Cressie estimator to be less influenced by outliers than the Matheron estimator the effects of outliers are still noteworthy.  

Our particular interest in this work is in spatially aggregated outliers that occur not isolated but in blocks of several close-by outlying observations. Block outliers can be considered as spatial change-points in the data, which have been considered recently in other contexts \citep{Otto2016, Garthoff2017}. An example of such outliers is clouds in satellite data. Consider for example data collected by the Landsat 8 satellite belonging to the NASA Landsat project. This satellite collects data of the earth’s land surface on 9 different spectral bands \citep{Knight2014}. Such data are strongly affected by weather effects. For example, clouds lead to measurement problems across possibly large regions and therefore to contamination in the form of outlier blocks. A further example of data with an outlier block is a soil survey with a leaking pipeline leading to contamination in an area around the leak. \citet{Kerry2007a} investigated the Matheron estimator in the case of spatially aggregated outliers but did not consider the Genton %and the Dowd 
estimator in this respect. In our simulations, reported in Section \ref{ch:block}, we find that this estimator is strongly influenced by block outliers.

A common assumption for random fields, usually made for convenience, is isotropy \citep[p. 87]{Sherman2011}. This means that the spatial dependency between two locations $\boldsymbol{s}, t \in I$ only depends on the Euclidean distance between $\boldsymbol{s}$ and $t$, but not on the direction. As mentioned in \citet[p. 87 ff.]{Sherman2011} falsely assuming isotropy can lead to some undesirable effects in the analysis. The assumption of isotropy is not reasonable or at least questionable in many applications. Considering, for example, crop yields, it is reasonable that the dependency structure depends on the wind direction, since the wind is important for the transportation of the pollen \citep{Guan2004}. Estimation of the dependency structure for different directions and therefore estimation of the variogram in different directions is necessary to check for isotropy or to consider anisotropy in the analysis. We focus here on the variogram estimation for anisotropic random fields. 

We suggest two directional variogram estimators for data observed on a regular grid. Both estimators are based on the highly robust Minimum Covariance Determinant (MCD) estimator of multivariate location and scatter introduced by \citet{Rousseeuw1985} and achieve robustness especially against block outliers. The MCD has already been used in time series analysis to construct a robust autocorrelation estimator with good performance in the presence of outlier patches \citep{Durre2015}. We use the MCD to construct  variogram estimators which jointly estimate the variogram at several spatial lags. Two different constructions are investigated here. The first one is based on vectors of subsequent observations  and estimates the variance and the covariances for different lags using the MCD. 
In the second approach vectors of pairwise differences of the spatial process are built, so that the variogram is directly captured on the main diagonal of the variance-covariance matrix of the vectors. We investigate the efficiency and the robustness of these and other proposals in worst case and in block outlier configurations in the following.
	
The rest of the paper is organized as follows. The MCD estimator is explained in Section 2. Section 3 describes the MCD based variogram estimators constructed here. We evaluate the estimators theoretically and via simulation in Section 4. Section 5 contains an application to data taken by the Landsat 8 satellite. Section 6 provides a summary and an outlook. 

%%%%%%%%%%%%%%%%%%%%%%%%%%
\section{MCD Estimator}\label{ch:MCD}
The Minimum Covariance Determinant (MCD) estimator is an affine equivariant and highly robust estimator of multivariate location and scatter, which was introduced by \citet{Rousseeuw1985}. The idea is to search for the $k(n)$ data points leading to the confidence ellipsoid with the smallest covariance determinant. The estimated location is the center of this ellipsoid and the estimated covariance matrix reflects its shape. 
Given $p-$dimensional data points $\boldsymbol{x}_1, \ldots , \boldsymbol{x}_n$, the (raw) MCD location and scatter estimator using a tuning constant $k(n)$ with $\lfloor\nicefrac{(n+p+1)}{2}\rfloor \leq k(n) \leq n$ is defined as follows \citep{Hubert2010}:
\begin{itemize}
	\item[1.] Search the $k(n)$ observations for which the sample covariance matrix has minimal determinant.
	\item[2.] Define $\widehat{\boldsymbol{\mu}}$ to be the mean of these $k$ observations.
	\item[3.] Set $\widehat{\boldsymbol{\Sigma}}$ to the sample covariance matrix of these observations multiplied by a consistency factor $c$.
\end{itemize}
The consistency factor depends on the distribution of the data, on the dimension of the vectors, and on the ratio of $k(n)$ to $n$ and  ensures that the estimator is Fisher-consistent. For independent identically normally distributed data the consistency factor is $c =  \nicefrac{\alpha}{ F_{\chi^2_{p+2}}(\chi^2_{p, \alpha})}$ asymptotically, whereby $\alpha = \lim_{n\rightarrow \infty} \nicefrac{k(n)}{n}$, $F_{\chi^2_{q}}$ is the distribution function and $\chi^2_{q,\alpha}$ the $\alpha$-quantile of the $\chi^2-$ distribution with $q$ degrees of freedom \citep{Croux1999}.  \citet{Pison2002} determine additional finite-sample correction factors to correct bias in small samples.
	
The MCD estimator of location and scatter is consistent and asymptotically normally distributed in case of elliptical distributions \citep{Butler1993}.  
It is highly robust with a finite sample breakdown point 
of $\left[(n-p+2)/2\right]/n$, which is
$0.5$ asymptotically,  if  $k(n) = \left[\nicefrac{(n+p+1)}{2}\right]$ and the sample is in general position \citep{Roelant2009}.  Accordingly, almost half of the observations can be replaced by arbitrary values before the estimator can exceed all bounds. Unfortunately, this estimator has rather low efficiency in case of an underlying normal distribution if $k(n)$ is chosen small to achieve high robustness. A reweighting step is often included to increase the efficiency of the MCD while retaining its robustness widely. The reweighted MCD is defined as 
	\begin{align*}
	\widehat{\boldsymbol{\mu}} &= \frac{\sum_{i=1}^n W(d_i^2)\boldsymbol{x}_i}{\sum_{i=1}^n W(d_i^2)} \\
		\widehat{\boldsymbol{\Sigma}} &= c_* \frac{1}{n}\sum_{i=1}^nW(d_i^2)\left(\boldsymbol{x}_i - \widehat{\boldsymbol{\mu}}\right)\left(\boldsymbol{x}_i-\widehat{\boldsymbol{\mu}}\right)^T
	\end{align*}
with $d_i^2 = \left(\boldsymbol{x}_i-\widehat{\boldsymbol{\mu}}_0\right)^T\widehat{\boldsymbol{\Sigma}}_0^{-1}\left(\boldsymbol{x}_i-\widehat{\boldsymbol{\mu}}_0\right)$, $W(\cdot)$ a weight function, $\widehat{\boldsymbol{\mu}}_0, \widehat{\boldsymbol{\Sigma}}_0$ the initial MCD estimation, and $c_*$ a consistency factor. A common choice for the weight function is $W(d^2) = \mathbb{I}\left(d^2\leq\chi^2_{p,0.975}\right)$ with $\mathbb{I}(\cdot)$ denoting an indicator function. The reweighting step improves the efficiency of the estimator while retaining the breakdown point, if the weight function is bounded and becomes zero for large distances $d_i$ \citep{Lopuhaa1991, Lopuhaa1999}.
	
Exact computation of the MCD estimator requires evaluation of all ${n \choose k(n)}$  subsets of size $k(n)$  \citep{Hubert2010}. In our simulations we use the FAST-MCD algorithm suggested by \citet{Rousseeuw1999}, which reduces the computation time a lot but does not necessarily lead to the global optimum. Further algorithms have been suggested by \citet{Hawkins2002},  \citet{Hubert2012} and \citet{Todorov1992}.

\section{Directional MCD variogram estimator}\label{ch:mcdvar}
Now we use the MCD estimator to estimate the variogram at several lags jointly. We consider two different approaches to construct directional variogram estimations based on the MCD. In the first approach vectors of pairwise differences are built such that the variogram is directly captured on the main diagonal of the variance-covariance matrix (Section 3.1). In the second approach vectors of the data at different locations are built such that the variogram can be calculated from the estimated variance-covariance matrix (Section 3.2). 
	
The MCD variogram estimators are constructed for directional variogram estimation in the four directions south-north (S-N), east-west (E-W), southwest-northeast (SW-NE) and southeast-northwest (SE-NW) under the assumption of observing a random field $\{Z(\boldsymbol{s}),~\boldsymbol{s}\in I\}$ on a two-dimensional regular grid $I = \{1, \ldots , n_x\} \times \{1, \ldots , n_y\}$ of size $n = n_x \times n_y$. An example for data observed on a regular grid are satellite data. In the following both MCD variogram estimators are explained for estimation in the E-W direction (along the x-axis). Estimation in the other directions works analogously and is briefly explained at the end of both subsections.

\subsection{MCD variogram estimator based on differences}\label{ch:mcddiff}
	
Now consider an intrinsically stationary and mean stationary random field $Z(\boldsymbol{s})$. For the MCD variogram estimator based on differences (MCD.diff)  vectors of differences between the observations are built such that the main diagonal of the estimated variance-covariance matrix captures the variogram up to the lag vector $\boldsymbol{h}_{h_{\max}}$. 
More precisely, vectors with the following structure are built
	\begin{align}
		\boldsymbol{W} = \begin{pmatrix}
			Z(\boldsymbol{s}) - Z(\boldsymbol{s}+\boldsymbol{h}_1) \\
			Z(\boldsymbol{s}) - Z(\boldsymbol{s}+\boldsymbol{h}_2) \\
			\vdots \\
			Z(\boldsymbol{s}) - Z(\boldsymbol{s}+\boldsymbol{h}_{h_{\max}})
		\end{pmatrix}
	\end{align}
for lag vectors $\boldsymbol{h}_1, \ldots , \boldsymbol{h}_{h_{\max}}$ as described in Section 3.1. Due to the assumption of intrinsic stationarity all vectors with this form have the same covariance matrix. Thereby, the variance of the first component equals the variogram at lag $\boldsymbol{h}_1$, the variance of the second component the variogram at lag $\boldsymbol{h}_2$, and so on. 
	
The lag vectors  $\boldsymbol{h}_l = (l,0)^T$,  $l = 1, \ldots , h_{\max}$, for estimation in the E-W direction lead to the $n_W = n_y \cdot (n_x - h_{\max})$ vectors
	\begin{align*}
		\begin{pmatrix}
			Z((x_1,y_1)^T) - Z((x_2, y_1)^T) \\ \vdots \\ Z((x_1, y_1)^T) - Z((x_{h_{\max}+1}, y_1)^T)
		\end{pmatrix},  \ldots, 
		\begin{pmatrix}
			Z((x_{n_x-h_{\max}}, y_{n_y})^T) - Z((x_{n_x-h_{\max}+1}, y_{n_y})^T) \\ \vdots \\ Z((x_{n_x-h_{\max}}, y_{n_y})^T) - Z((x_{n_x},y_{n_y})^T)
		\end{pmatrix}
	\end{align*}
The first component is a pairwise difference of the process for two locations with an Euclidean distance of $||\boldsymbol{h}_1|| = 1$ in E-W direction,  the second component a pairwise difference for lag $||\boldsymbol{h}_2 || = 2$ in the same direction and so on. The variance-covariance matrix of these vectors is estimated using the MCD estimator. MCD.diff is then given by 
	\begin{align*}
		(2\widehat{\gamma}(\boldsymbol{h}_1), \ldots , 2\widehat{\gamma}(\boldsymbol{h}_{h_{\max}}))^T = \text{diag}\left(\widehat{\boldsymbol{\Sigma}}_{MCD}\right).
	\end{align*}
For estimation of the variogram in one of the other directions the vectors are adapted such that the $l$-th component of each vector is a pairwise difference for lag $\boldsymbol{h}_l,l=1,\ldots , h_{\max}$ in the desired direction.

\subsection{MCD variogram based on the original data}\label{ch:mcdorg}
	 
For the first approach, which is called MCD.org in the following, we need the assumption that $Z(\boldsymbol{s})$ is weakly stationary, i.e., that it is mean stationary and that the covariances are translation-invariant. This is somewhat stronger than the assumption of intrinsic stationarity assumed e.g. for the Matheron estimator. The $n = n_x \cdot n_y$ spatial coordinates are labeled with $\boldsymbol{s}_1 = (x_1, y_1)^T, \ldots , \boldsymbol{s}_n = (x_{n_x}, y_{n_y})^T$. Thereby let $x_i,~i=1,\ldots , n_x$, be the east-west coordinate (on the x-axis) and $y_j,~j=1,\ldots ,n_y$, the north-south coordinate (on the y-axis). 
	
Under the assumption of  weak stationarity we have the following relationship between covariances of the process and the variogram 
	\begin{align}
		2\gamma(\boldsymbol{h}) = 2[\text{Var}(Z(\boldsymbol{s})) - \text{Cov}(Z(\boldsymbol{s}), Z(\boldsymbol{s}+\boldsymbol{h}))].
		\label{form:vario-cov}
	\end{align}
Therefore, a variogram estimation can be derived from estimations of the variance and the covariance. If $h_{\max}$ is the number of lags to be estimated, $h_{\max}+1$ dimensional vectors with the following structure are built
	\begin{align}
		\boldsymbol{V} = \begin{pmatrix}
			Z(\boldsymbol{s}) \\ Z(\boldsymbol{s}+\boldsymbol{h}_1) \\ \vdots \\ Z(\boldsymbol{s}+\boldsymbol{h}_{h_{\max}})
		\end{pmatrix},
	\label{form:vec-MCDorg}
	\end{align} 
with $h_1, \ldots, h_{h_{\max}}$ being lag vectors in one of the four directions. The choice of $h_{\max}$ depends on the data and the aim of the analysis. However, $h_{\max}$ should not be larger than necessary, because the higher $h_{\max}$ the fewer vectors are available for the estimation. We consider equidistant choices $\boldsymbol{h}_1, \ldots, \boldsymbol{h}_{h_{\max}}$ here, but other choices are also possible. For estimation in the E-W direction e.g. we use the vectors $\boldsymbol{h}_{l} = (l,0)^T, l = 1, \ldots ,h_{\max}$. In this case, $h_{\max} = ||\boldsymbol{h}_{h_{\max}}||$ holds. Due to the assumption of translation-invariant covariances, these vectors have the variance-covariance matrix
	\begin{align}
		\boldsymbol{\Sigma}_{\boldsymbol{V}} = \begin{pmatrix}
			a_0 & a_1 & a_2 & \ldots & a_{h_{\max}} \\
			a_1 & a_0 & \ddots & \ddots & \vdots \\
			a_2 & \ddots & \ddots & \ddots & a_2    \\
			\vdots & \ddots & \ddots & \ddots & a_1 \\
			a_{h_{\max}} & \ldots & a_2 & a_1 & a_0
		\end{pmatrix} 
	\in \mathbb{R}^{(h_{\max} + 1) \times (h_{\max}+1)}
	\label{form:cov-MCDorg}
	\end{align}
with $a_0 = \text{Var}(Z(\boldsymbol{s}))$, $a_1 = \text{Cov}(Z(\boldsymbol{s}), Z(\boldsymbol{s} +\boldsymbol{h}_1))$, $a_2 = \text{Cov}(Z(\boldsymbol{s}), Z(\boldsymbol{s}+\boldsymbol{h}_2))$ and so on. 
We apply the raw or the reweighted MCD estimator to estimate the variances and covariances in this matrix and use formula (\ref{form:vario-cov}) to obtain the variogram estimation.

For estimation in E-W direction we have the following $n_V = n_y \cdot (n_x - h_{\max})$ vectors available:
	\begin{align*}
		\begin{pmatrix}
			Z((x_1,y_1)^T) \\ \vdots \\ Z((x_{h_{\max}+1}, y_1)^T)
		\end{pmatrix}, \ldots, 
		\begin{pmatrix}
			Z((x_{n_x-h_{\max}},y_1)^T) \\ \vdots \\ Z((x_{n_{x}}, y_1)^T)
		\end{pmatrix}, \ldots , 
		\begin{pmatrix}
			Z((x_{n_x-h_{\max}}, y_{n_y})^T) \\ \vdots \\ Z((x_{n_{x}}, y_{n_y})^T)
		\end{pmatrix}
	\end{align*}
Thus, for each $y$-coordinate in north-south direction we build vectors with $h_{\max}+1$ subsequent observations in the E-W direction. 
The MCD estimator is then used to estimate the variance-covariance matrix of the vectors. As seen in formula (\ref{form:cov-MCDorg}), several different estimations for the variance of the process are obtained on the main diagonal and several estimations for the covariance at lag $||\boldsymbol{h}_l||$ on the $l$-th minor diagonal. By averaging the different estimates and using formula (\ref{form:vario-cov}) we obtain estimations $\widehat{\gamma}(\boldsymbol{h}_l),~l=1,\ldots , h_{\max}$. 
		
Directional variogram estimations in S-N, SW-NE or SE-NW direction can be calculated analogously. For this we build vectors with the structure of formula (\ref{form:vec-MCDorg}) with $\boldsymbol{h}_1, \ldots , \boldsymbol{h}_{h_{\max}}$ being lag vectors in the direction of interest. For  estimation in north-south direction, e.g.,  vectors of subsequent observations in north-south direction are constructed for each west-east coordinate. For the estimation in SW-NE $\boldsymbol{h}_l = (l,l)^T,l=1, \ldots , h_{\max}$ and for the estimation in SE-NW $\boldsymbol{h}_l = (l,-l)^T, l=1,\ldots , h_{\max}$. Note, in these two cases $h_{\max} \neq ||h_{h_{\max}}||$.
	
%%%%%%%%%%%%%%%%%
 \section{Evaluation of the estimators}\label{ch:sim}
 
In the following we evaluate characteristics of the estimators such as efficiency, consistency and robustness as measured e.g. by breakdown points. Theoretical derivations are difficult due to the spatial dependency of the data and the complexity of the MCD estimator. Therefore we use some simplifications in the theoretical considerations and additional simulations to  investigate the behavior of the estimators and compare them to the common Matheron and Genton variogram estimators.

\begin{figure}[H]
		\center
		\includegraphics[scale = 0.5]{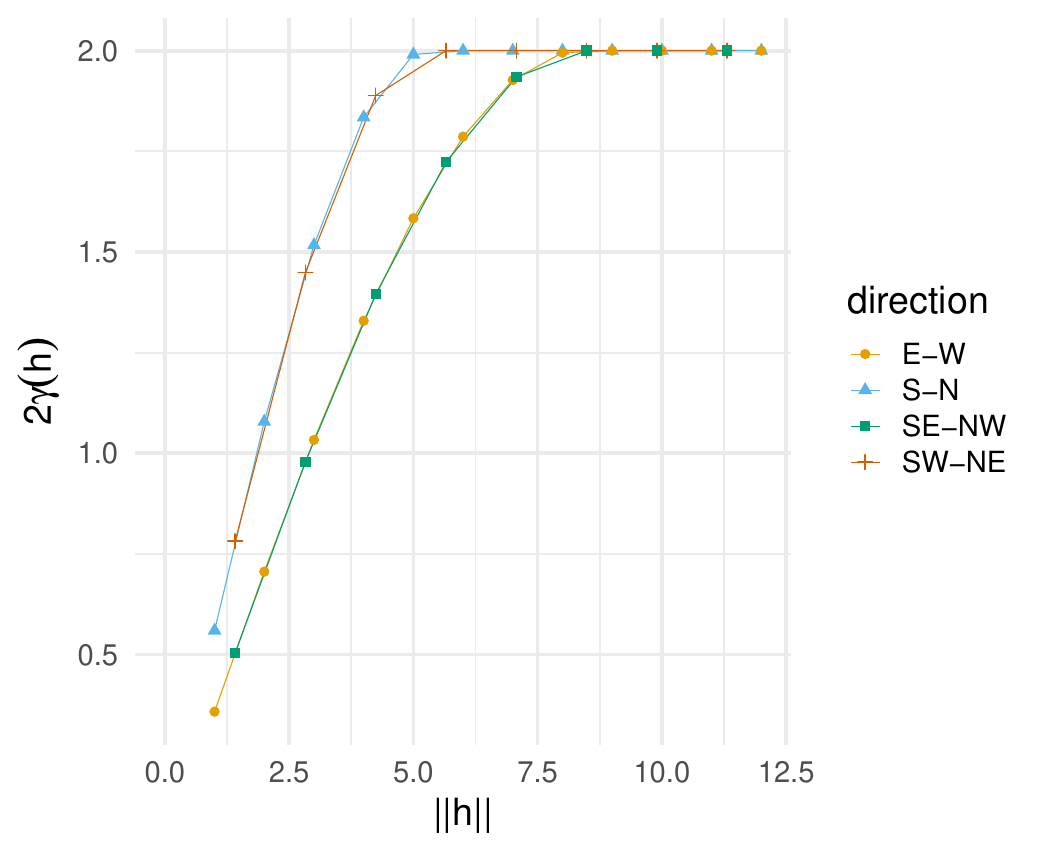}
		\caption{Geometric anisotropic spherical variogram for different directions used for the generation of the data}
		\label{fig:sph_true}
\end{figure}

In the simulations we consider a Gaussian random field $Z(\boldsymbol{s})$ on a two-dimensional regular grid of size $15 \times 15$ corresponding to a total sample size of $n=225$ observations with a geometric anisotropic variogram. Geometric anisotropy means that the dependence structure can be converted to an isotropic one through rotation with angle and rescaling of the coordinate axes. The rotation matrix $R$, the rescaling matrix $T$ and the variogram are given by 
\begin{align*}
    \boldsymbol{R} = \begin{pmatrix}
        \cos(\theta) & \sin(\theta) \\
        -\sin(\theta) & \cos(\theta)
    \end{pmatrix},~~
    \boldsymbol{T} = \begin{pmatrix}
        1 & 0 \\
        0 & \frac{1}{b}^{\frac{1}{2}}
    \end{pmatrix}, ~~
    2\gamma(\boldsymbol{h}) = 2\gamma_{0}\left(\sqrt{\boldsymbol{h}^{T}\boldsymbol{R}^{T}\boldsymbol{T}^{T}\boldsymbol{T}\boldsymbol{R}\boldsymbol{h}}\right)
\end{align*}
with $2\gamma_0(\cdot)$ an isotropic variogram \citep[pp. 91]{Sherman2011}. \\
For the isotropic variogram $2\gamma_0(||\boldsymbol{h}||)$ we choose a spherical variogram with a range $R=5$ and a sill $\beta=2$. The rotation angle $\theta$ is $\nicefrac{3\cdot\pi}{8}$ and the ratio $b$ is equal to $2$. The formula of the spherical variogram is 
	\begin{align*}
	    2\gamma_{0}(||\boldsymbol{h}||) =
	    \beta \left( \frac{3||\boldsymbol{h}||}{2R}-\frac{||\boldsymbol{h}||^3}{2R^3}\right)   	    \mathbb{I}(0 < ||\boldsymbol{h}|| < R)+
	    \beta \mathbb{I}(||\boldsymbol{h}|| \geq R).
	\end{align*} 
The variogram for the different directions is illustrated in Figure \ref{fig:sph_true}. For the directions E-W and S-N lags up to $h_{\max}=7$, with $||\boldsymbol{h}_7|| = 7$, are chosen. For the SW-NE and SE-NW direction we set $h_{max}=5$ and use, e.g.,  the vectors $\boldsymbol{h}_i = (i,i)^T$, $i=1, \ldots , 5$ with $||\boldsymbol{h}_5||=7.07$ for the SW-NE direction. Therefore, for the E-W and SE-NW direction the largest estimated lag is smaller than the range of the true variogram, while for the S-N and SW-NE direction the largest estimated lag is higher than the range.
In all simulations 1000 samples are used for each setting. 
Our aim is the estimation of the variogram $\gamma(\boldsymbol{h})$ of the process $Z(\boldsymbol{s})$.

The simulations are performed on a HPC cluster using the software \texttt{R} 4.4.0 \citep{r2024}. The R-package \texttt{RandomFields} \citep{randomfiels2020} is used for the simulation of the Gaussian random field. The MCD estimator is calculated using \texttt{robustbase} \citep{robustbase2021}. All values are rounded to $2$ decimal places. MCD.org.re and MCD.diff.re are in the following the MCD variogram estimators using the reweighted version of the MCD estimator. In further simulations not presented here, we considered further  grid sizes, other variogram models and different values of $h_{\max}$. The results are qualitatively similar to the results presented in the following.  
In graphical illustrations, we connect the point estimations for different lags by a line to improve the visualization.

\subsection{Correction factors}\label{ch:corrfac}
The MCD scale estimation needs to be multiplied with a consistency factor, which depends on the distribution of the data and the dimension of the vectors \citep{Croux1999}. For small sample sizes \citet{Pison2002} derive an additional finite-sample correction factor for the MCD scale, which depends on the sample size and correct the bias for small samples. 
Due to the spatial dependency of the data theoretical derivation of a consistency factor for MCD.org and MCD.diff are difficult. Therefore, we use the consistency factor of the MCD estimator for normally distributed data here. In simulations we see that these correction factors work well also in our context, particularly if the size of the grid is large.  

We derive additional finite-sample bias correction factors for variogram estimation by MCD.org and MCD.diff using simulations for further improvement, particularly in case of small grids. These factors are to be applied in addition to the asymptotic correction factors. For the sake of fairness in the following simulations, we also determine such correction factors for the Matheron and the Genton estimator.
For this we generate 1000 samples for each of several scenarios. Let $\widehat{\gamma}_l(\boldsymbol{h}_i)$ be the semivariogram estimation for lag $\boldsymbol{h}_i,~i=1,\ldots , h_{\max}$ and dataset $l=1,\ldots , 1000$. Then our simulated correction factors are obtained as
	\begin{align}
		c_{\text{opt}} = \left(\frac{1}{h_{\max}} \sum_{i=1}^{h_{\max}-1}\frac{1}{1000}\sum_{l=1}^{1000}\frac{\widehat{\gamma}_l(\boldsymbol{h}_i)}{\gamma(\boldsymbol{h}_i)}\right)^{-1}.
		\label{form:corrfac}
	\end{align}
We only average over all lags except the largest one because all robust estimators investigated here (including the one of Genton) reflect the correct shape of the true variogram for small lags but not for large lags. Especially for the largest lag the estimators generally underestimate the true variogram. Averaging across all lags would thus lead to worse results. 

The simulated finite sample correction factors are reported for all estimators and all four directions in Table \ref{tab:corrfac}.
For the raw MCD estimators (MCD.org and MCD.diff) the simulated factors for the  E-W and S-N direction are similar and a little higher than those for the SW-NE and SE-NW direction. Due to the rotational symmetry of the grid the same number of vectors can be built for estimation in E-W and S-N direction, which is larger than that for estimation in SW-NE and SE-NW direction (see Table \ref{tab:amount_vec} in the appendix). The correction factors for these estimators seem to depend on the number of vectors. The simulated values for the reweighted versions  MCD.org.re and MCD.diff.re are similar for all four directions. They are higher than those for the raw estimators and in particular, they are larger than 1. The reweighted MCD estimators underestimate the variogram, whereas the raw MCD estimators overestimate. 

In Table \ref{tab:corrfac_2} in the appendix the simulated correction factors for two other variogram models are shown. The left half is for an exponential variogram model and the right half for a Gaussian variogram model. We see that especially for the reweighted estimators the correction factors depends on the unknown true variogram model, and this also applies to the Genton estimator. For the raw MCD estimators the
 influence of the unknown true model seems to be small. 
Further simulations not reported here indicate that the correction factor is mostly determined by the number of vectors available for the estimation, which depends on the grid size, by $h_{\max}$ and the direction. The more vectors can be built, the closer the additional finite-sample correction factor gets to 1. For large grids like a $50\times 50$ grid the simulated correction factors for all estimators are almost equal to one and can thus be ignored, irrespective of  the true variogram model. \\
Note that these additional correction factors are not far from 1 especially for the raw MCD estimators, i.e., neglecting them just results in a small positive bias.  
	 
\begin{table}
		\centering
		\begin{tabular}{l|rrrr}
			& \multicolumn{4}{|c}{direction} \\ 
			estimator & S-N & E-W & SW-NE & SE-NW \\ \hline
            Matheron & 1.00 & 1.00 & 1.00 & 1.00 \\ 
            Genton & 1.03 & 1.03 & 1.02 & 1.03 \\ 
            MCD.diff & 0.96 & 0.97 & 0.90 & 0.91 \\ 
            MCD.diff.re & 1.19 & 1.17 & 1.19 & 1.16 \\ 
            MCD.org & 0.95 & 0.96 & 0.88 & 0.91 \\ 
            MCD.org.re & 1.20 & 1.18 & 1.19 & 1.19 \\  
            \hline
		\end{tabular}
	\caption{Simulated finite sample correction factors for the variogram estimators in all four directions in case of a regular 15$\times$ 15 grid}
	\label{tab:corrfac}
\end{table}

%%%%%%%%%%%%%	
\subsection{Gaussian data}\label{ch:oA}

\begin{figure}[H]
        \centering
		\includegraphics[scale = 0.8]{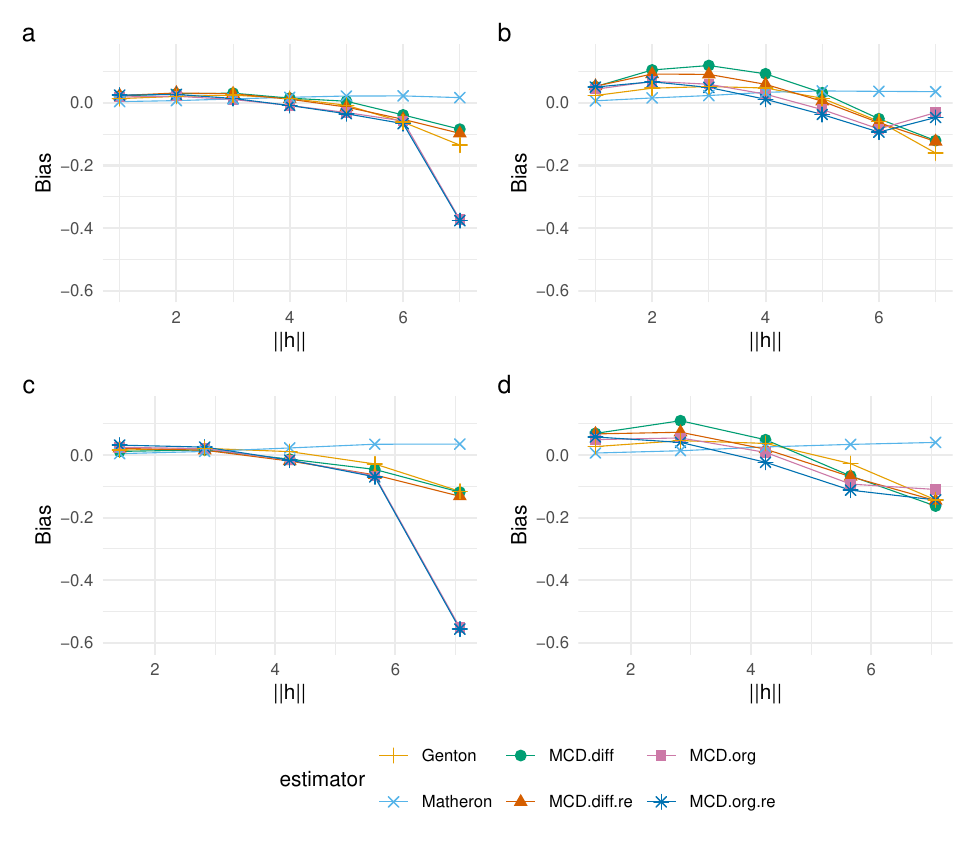}
			\caption{Estimated bias of the variogram estimation in E-W (a), S-N (b), SE-NW (c) and SW-NE (d) direction by several estimators}
		\label{fig:Bias_oA}
\end{figure}

\begin{figure}[H]
    \centering
	\includegraphics[scale = 0.8]{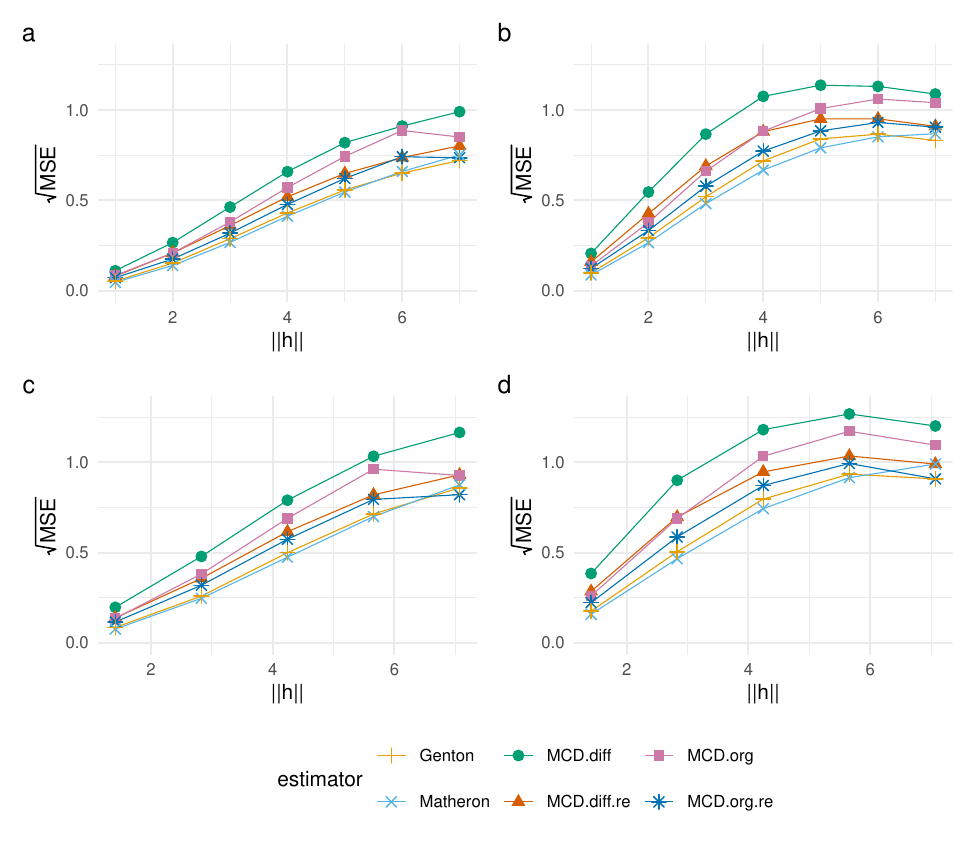}
	\caption{Root mean squared errors of the variogram estimations in E-W (a), S-N (b), SE-NW (c) and SW-NE (d) direction}
	\label{fig:MSE_oA}
\end{figure}

We first consider simulation results for non-contaminated Gaussian random fields as described at the beginning of Section 4.   
Figure \ref{fig:Bias_oA} shows the estimated bias and Figure \ref{fig:MSE_oA} the root of the mean square error rMSE for the estimation in all four directions of the Matheron estimator, the Genton estimator, the raw MCD estimators (MCD.org and MCD.diff) and the reweighted MCD estimator (MCD.org.re and MCD.diff.re).  
In Figure \ref{fig:sph_true} we see that the variograms for the E-W and the SE-NW direction are almost identical, and the same applies to the variograms for the S-N and SW-NE direction. This also applies to the bias. The estimated bias and rMSE for the E-W and SE-NW direction (Figure \ref{fig:Bias_oA} a, c and \ref{fig:MSE_oA} a, c) are similar, just like the estimated bias for the S-N and SW-NE direction (Figure \ref{fig:Bias_oA} b, d and \ref{fig:MSE_oA} b, d). 
This matches the true variograms for the different directions (see Figure \ref{fig:sph_true}), because the true variogram of the S-N and SW-NE direction are the same, just like the true variogram for the E-W and the SE-NW direction.

Figure \ref{fig:Bias_oA} shows that the Matheron estimator is nearly unbiased for all lags and all directions, while the other estimators show  some bias and underestimate for lager lags. Considering first the E-W and SE-NW direction, we see that the Genton estimator is nearly unbiased for small lags and underestimates the true variogram for lager lags. The raw and the reweighted MCD.diff estimator in these two directions  have a bias similar to that of the Genton estimator. In contrast, MCD.org and MCD.org.re are also nearly unbiased for small lags but strongly underestimate for the largest lag for estimation in these two directions. It seems that these two estimators have a problem with estimating the largest lag if $h_{\max}$ is smaller than the range of the true variogram. 
For the other two directions S-N and SW-NE all estimators except the Matheron estimator overestimate the variogram slightly for small lags and underestimate for lager lags. Especially for the larger lags the Genton estimator shows smaller bias than the MCD estimators but the differences are small. 
In general, the bias of the raw MCD estimators (MCD.org and MCD.diff) and the reweighted MCD estimators (MCD.org.re and MCD.diff.re) are very similar. 

Figure \ref{fig:MSE_oA} illustrates that the Matheron estimator is the most efficient estimator in terms of the rMSE in case of  observations from a Gaussian random field without outliers. The rMSE of the Genton estimator is close to that of the Matheron estimator. The MCD based estimators are less efficient than these, but the reweighting reduces this loss  substantially. MCD.org turns out to be more efficient than MCD.diff in all situations considered here.
	
%%%%%%%%%%%%%%%%%%
\subsection{Consistency}\label{ch:cons}

The consistency of  MCD based variogram estimation can be argued analytically under some assumptions concerning the dependence structure and some modification. In this subsection we assume $m_y-$dependence in the north-south direction and $m_x-$dependence in the east-west direction  for some positive integers $m_x$ and $m_y$, and we consider variogram estimation in the latter direction.  $Z((x,y)^T)$ and $Z((x+h,y)^T)$ are independent then for $||\boldsymbol{h}||> m_x$. Modified versions of the two MCD variogram estimators, which skip $m_y$ rows and use only non-overlapping vectors with a minimal difference of $m_x+1$ between the components of different vectors in E-W direction, can easily be seen to be consistent as $n_x\cdot n_y$ increases to infinity  because of the consistency of the MCD when applied to independent vectors \citep{Butler1993}. 

Such modified estimators use only a small part of the available observations and of the possible vectors we can built. The efficiency can be increased by averaging the results across  all the  different  possibilities to choose independent vectors for such modified MCD variograms. For example, for estimation in E-W direction we can build non-overlapping vectors with a minimal distance of $m_x+1$ starting the first vector in each row with $Z((1,y)^T)$, or we can shift all vectors starting the first vector with $Z((2,y)^T)$, etc., and similarly we can select different subsets of independent rows from the grid. The average of the resulting different modified MCD variogram estimations will again be consistent in such scenarios as the mean of consistent estimators is also consistent. Thereby, a set of independent vectors is only used if it consists of more than $2\cdot h_{\max}$ vectors to avoid singularity problems with the MCD estimator. A drawback of this approach are that knowledge of suitable values $m_x$ and $m_y$ is needed. Another more relevant drawback is reduced robustness as compared to the estimators defined in Section 3, as we will see in Table \ref{tab:mA} in Subsection \ref{ch:block}. We nevertheless consider this (averaged and) modified MCD based variance estimator as a theoretically more tractable and consistent benchmark in the following.  

We investigate the behavior of all estimators including the average modified versions (name affix ".mod") for increasing grid sizes and the isotropic Gaussian process described at the beginning of Section 4. The average modified versions need knowledge of $m_x$ and $m_y$. Here we use the true values $m_x= 8$ and $m_y = 5$. The different grid sizes and the corresponding number of vectors used in the multivariate estimations can be found in Table \ref{tab:amount_vec} in the appendix.

\begin{figure}[H]
    \center
	\includegraphics[scale = 0.6]{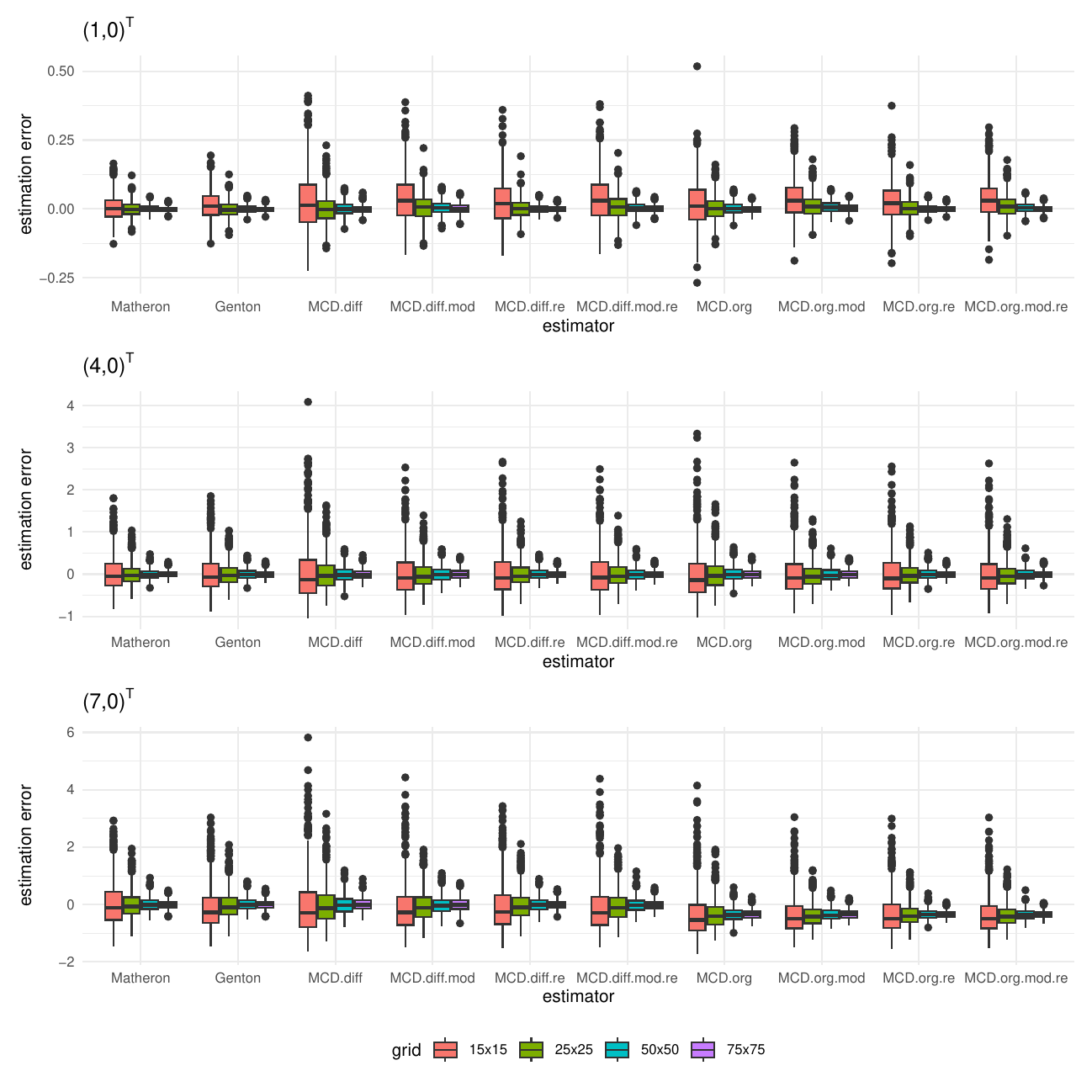}
	\caption{Boxplots of the estimation errors obtained by several estimators including the average modified versions in case of different grid sizes and three different lag vectors for the estimation in E-W direction in case of data from a Gaussian model}
\label{fig:Bias_boxplot_EW}
\end{figure}

Figure \ref{fig:Bias_boxplot_EW} shows boxplots of the resulting estimation errors for the E-W direction and three different lags. The variances of these errors are larger for higher lags, but all the variances decrease as the grid size increases. For small grid sizes, the Matheron and the Genton estimators have much smaller variances than those based on the MCD. Concerning bias, the Matheron estimator shows the best behavior in general. 
The average modified MCD estimators (name affix ".mod") show especially for small grid sizes a smaller variance than MCD.org and MCD.diff. The multivariate estimators based on the original data (raw and reweighted MCD.org and MCD.org.mod) have particularly  for small grid sizes smaller variance than the multivariate estimators based on differences (raw and reweighted MCD.diff and MCD.diff.mod). Reweighting (name affix ".re") improves the efficiency of the multivariate estimators introduced in Section \ref{ch:mcdvar} as it reduces the variance without increasing the bias.
The estimators get approximately unbiased as the grid size increases, except for the raw and the reweighted MCD.org and MCD.org.mod which  underestimate the true variogram at the largest lag for all grid sizes. In E-W direction $||\boldsymbol{h}_{h_{\max}}||$ is smaller than the range of the true variogram. As already seen in subsection \ref{ch:oA}, the multivariate estimator based on the original data has problems with estimating the largest lag in such cases. 

\begin{figure}[H]
	\includegraphics[scale=0.6]{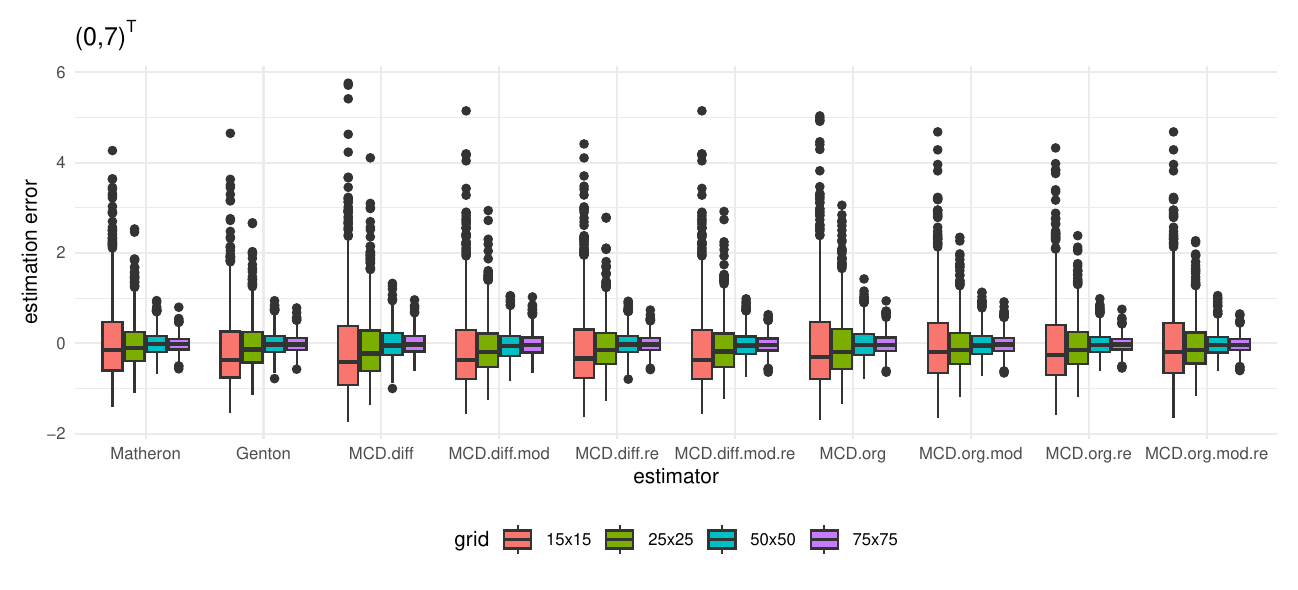}
	\caption{Boxplot of the estimation errors in S-N direction for several estimators including the average modified versions for different grid sizes and the lag vector $\boldsymbol{h}= (0,7)^T$}
	\label{fig:Bias_boxplot_SN}
\end{figure}

Figure \ref{fig:Bias_boxplot_SN} shows boxplots of the estimation errors for the largest lag vector $(0,7)^T$ in S-N direction. In this case $||\boldsymbol{h}_{h_{\max}}||$ is larger than the range of the true variogram. For this case the multivariate MCD estimators based on the original data are also unbiased for large grid sizes.
As the true range of the variogram is unknown it might be beneficial to ignore the largest estimated lag for MCD.org and MCD.org.re.

%%%

\subsection{Contamination with block outliers}\label{ch:block}

Now we investigate the robustness of the estimators against outliers, starting with the breakdown point of the modified MCD.org and MCD.diff estimators. We consider directional variogram estimation, building the vectors e.g. row by row for estimation in E-W direction. For the reason of simplicity we look at the  one-dimensional case $n_y = 1$, so that  the number of data points is $n=n_x$. An outlier block then corresponds to a sequence of subsequent outliers, which  could be caused by a temporary level shift or by measurement problems  resulting in a stretch of data with another level or a higher variance. We investigate the worst case to find the finite sample explosion breakdown point. Therefore, we determine the minimal length of an outlier block which can cause arbitrarily large values of the estimator. 

We start with the modified version of the MCD based estimators introduced in Subsection \ref{ch:cons}. Assuming an $m$-dependent process we build vectors with a minimal distance of $m+1$ between subsequent vectors.  We can build up to $n^{*}  = \lfloor \frac{n_x-h_{\max} – 1}{h_{\max}+1+m} +1 \rfloor$ non-overlapping vectors with $h_{\max}$ being the maximal lag we want to estimate. Here and in the following we assume these vectors to be in general position, so that the MCD scatter estimator with $k = \lfloor \frac{n+p+1}{2}\rfloor$ has the maximal finite sample breakdown point of $\varepsilon_{\text{MCD}} = \frac{n-k+1}{n}$. Thereby,  $n$ is the number of vectors, $k$ the size of the subsets considered by the MCD estimator and $p$ the dimension of the vectors \citep{Roelant2009}. Here, $n=n^{*}$ and $p = h_{\max}+1$ if the modified MCD.org is used and $p = h_{\max}$ if the modified MCD.diff is used. The number $\ell^{*}$ of disturbed vectors needs to be at least   $\lceil\varepsilon_{\text{MCD}}\cdot n^{*}\rceil = n^*-\lfloor \frac{n^{*}+p+1}{2}\rfloor+1$ to carry the estimator beyond all bounds. To disturb $\ell>1$ vectors the block must at least have a length of $l_{\ell} = 1 + (\ell-2)(m+h_{\max}+1)+m+1$ because the vectors have a distance of $m$. The minimal block length disturbing $\ell^{*}$ vectors is determined by $l_{\ell^*} = l_{n^*-\lfloor \frac{n^{*}+p+1}{2}\rfloor+1}$. This leads to the following breakdown point due to one block,
\begin{align*}
    \varepsilon_{\text{mod}}^{\text{Block}} = \frac{l_{n^*-\lfloor \frac{n^{*}+p+1}{2}\rfloor+1}}{n_x} = \frac{1 + (n^{*}-\lfloor \frac{n^{*}+p+1}{2}\rfloor- 1)(m+h_{\max}+1)+m+1}{n_x},
\end{align*}
which depends on $h_{\max}$, $n_x$, $m$ and $p$. The larger $h_{\max}$ and $m$, the smaller $n^{*}$ and the breakdown point.  If $h_{\max}$ and $m$ are large and $n_x$ is rather small, the lower bound of $\varepsilon_{\text{mod}}^{\text{Block}}$ becomes small. In case of $n_x = 50, m = 1$ and $h_{\max} = 4$, we get $n^{*} = 8$. This leads to the breakdown point  $\varepsilon_{\text{mod}}^{\text{Block}} = \frac{3}{50} = 0.06$ (MCD.org.mod) and $\varepsilon_{\text{mod}}^{\text{Block}} = \frac{9}{50} = 0.18$ (MCD.diff.mod).
Furthermore we must choose $h_{\max}$ carefully to ensure that $n^{*}$ is larger than the dimension of the vectors $h_{\max}+1$. In case of $n_x = 50$, $m = 5$ and $h_{max}=6$ lags, we only can build 4 vectors for the modified MCD and cannot use this estimator. Averaging the estimates obtained from different partitions as suggested in Subsection \ref{ch:cons} does not change this value. 

Next we discuss explosion of MCD.org and MCD.diff in case of an outlier sequence in one-dimensional data. A sequence of $l$ outliers disturbs up to $h_{\max} + l$ vectors, and the number of vectors which can be built is $n^{*} = n_x - h_{\max}$.  Thus $h_{\max}+l$ needs to be at least equal to $\lceil\varepsilon_{\text{MCD}}\cdot n^{*}\rceil = n^*-\lfloor \frac{n^{*}+p+1}{2}\rfloor+1$ to break the estimator. This corresponds to a minimal sequence length $l_{\min}=n^*-\lfloor \frac{n^{*}+p+1}{2}\rfloor+1 - h_{\max}$ and the following explosion breakdown  point
due to an outlier block \begin{align*}
    \varepsilon_{\text{MCD.var}}^{\text{Block}} = \frac{l_{\min}}{n_x} = \frac{n^*-\lfloor \frac{n^{*}+p+1}{2}\rfloor+1 - h_{\max}}{n_x}.
\end{align*}
In case of $n_x = 50$ and $h_{\max} = 4$ we have $\varepsilon_{\text{MCD.var}}^{\text{Block}} = 0.34$ for MCD.org and $\varepsilon_{\text{MCD.var}}^{\text{Block}} = 0.36$ for MCD.diff, which is much higher than the bounds for the modified estimators derived above. 
The larger $h_{\max}$, the  smaller get these lower bounds for explosion breakdown.  

Finally, we consider breakdown of the Genton estimator for such scenarios, i.e. in the one dimensional case. This estimator treats each  lag separately. Let $h$ be the lag for which we want to estimate the variogram. An outlier block of length $l$ disturbs at most $l + min(l,h)$ differences. We can build $n^{\star} = n_x-h$ differences for the estimation at lag $h$. The Genton estimator applies the Qn estimator of \citet{Rousseeuw1993} with the finite sample explosion breakdown point $\varepsilon_{\text{Qn}} = \nicefrac{\lfloor\frac{n^{\star}+1}{2}\rfloor}{n^{\star}}$ to these differences. The minimal number of differences which needs to be contaminated for breakdown is $\lceil\varepsilon_{\text{Qn}} \cdot (n_x - h)\rceil$. The minimal length of an outlier block to break the estimator thus is $l_{\min} =\max\{ \lceil\varepsilon_{\text{Qn}} \cdot (n_x - h) \rceil- h, \lceil\varepsilon_{\text{Qn}} \cdot (n_x - h) \rceil/2\}$. This leads us to
\begin{align*}
\varepsilon_{\text{Genton}}^{\text{Block}}(h) = \frac{l_{\min}}{n_x} = \frac{\max\{ \lceil\varepsilon_{\text{Qn}} \cdot (n_x - h) \rceil- h, \lceil\varepsilon_{\text{Qn}} \cdot (n_x - h) \rceil/2\}}{n_x} \ .
\end{align*}
In case of $n_x = 50$ and $h = 4$ we have $\varepsilon_{\text{Qn}} = 0.5$, $l_{\min} = 19$ and therefore $\varepsilon_{\text{Genton}}^{\text{Block}}(4) = 0.38$, which is  similar to the breakdown points for MCD.org and MCD.diff.

These considerations  correspond to breakdown points in the one dimensional case. Moreover, the Qn underlying the Genton estimator is known to show a noteworthy bias already in case of contamination rates much smaller than its breakdown point. Now we  investigate the performance of the estimators in several 2-dimensional block outlier scenarios  via simulations. We consider the following statistical model with several contamination rates and contamination distributions with different expectations and variances. The process of interest is the Gaussian Random Field $Z(\boldsymbol{s})$ described at the beginning of Chapter 3. But instead of $Z(\boldsymbol{s})$ we observe a Process $Y(\boldsymbol{s})$. For each location $\boldsymbol{s}_0 = (x_0, y_0)^T$ we define a block neighborhood $N(\boldsymbol{s}_0)$ containing $\lceil \epsilon \cdot n \rceil$ locations, where $\epsilon$ the fraction of outliers. This neighborhood is selected to be as quadratic as possible with $\boldsymbol{s}_0$ being as central as possible. 
Furthermore, let $W\sim\mathcal{N}(\mu_0, \sigma_0^2)$ be a normally distributed random variable, which describes the outlier distribution, and $I$ a discrete random variable with a uniform distribution on all locations $\boldsymbol{s}_1, \ldots, \boldsymbol{s}_n$. $I$ selects one location $\boldsymbol{s}_0 = (x_0, y_0)^T$ randomly and  all locations in $N(\boldsymbol{s}_0)$ belong to a block of substitutive outliers.
The observed Process $Y(\boldsymbol{s})$ is defined as  
\begin{align*}
    Y(\boldsymbol{s})|(I=\boldsymbol{s}_0) = \begin{cases} 
    Z(\boldsymbol{s}),~\text{ if } \boldsymbol{s} \notin N(\boldsymbol{s}_0) \\
    W,~ \text{ if } \boldsymbol{s} \in N(\boldsymbol{s}_0).
    \end{cases}
\end{align*}
We positioned the outlier block randomly since simulations not reported here revealed that the position of the outlier block influences the performance of the estimators. We obtained similar results in case of additive instead of substitutive outliers. 

Table \ref{tab:mA} in the appendix reports the bias and the rMSE of the estimations in E-W and S-N direction for different contamination rates $\epsilon$ and different contamination distributions. The best result for each scenario and each lag is  underlined.  As expected, the higher the contamination rate and the more extreme the contamination distribution, the larger get the biases.  The biases of the Genton and even more the Matheron estimator increase more than those of the MCD based estimators, with the reweighted MCD.org delivering the best results regarding both bias and rMSE in most cases. Here, reweighting improves the variance without increasing the bias. However, the reweighted estimators have the disadvantage that the correction factors depend on the true variogram model for small grid sizes, whereas this dependence is much lower for the other estimators. The modified MCD estimators described in Section \ref{ch:cons} perform in some cases even worse than the Matheron estimator. The modified MCD estimators are in most cases less robust than the MCD estimators described in section \ref{ch:mcdvar}. Due to this inferior performance, the modified versions will not be investigated further.

\begin{figure}[H]
	\includegraphics[scale=0.8]{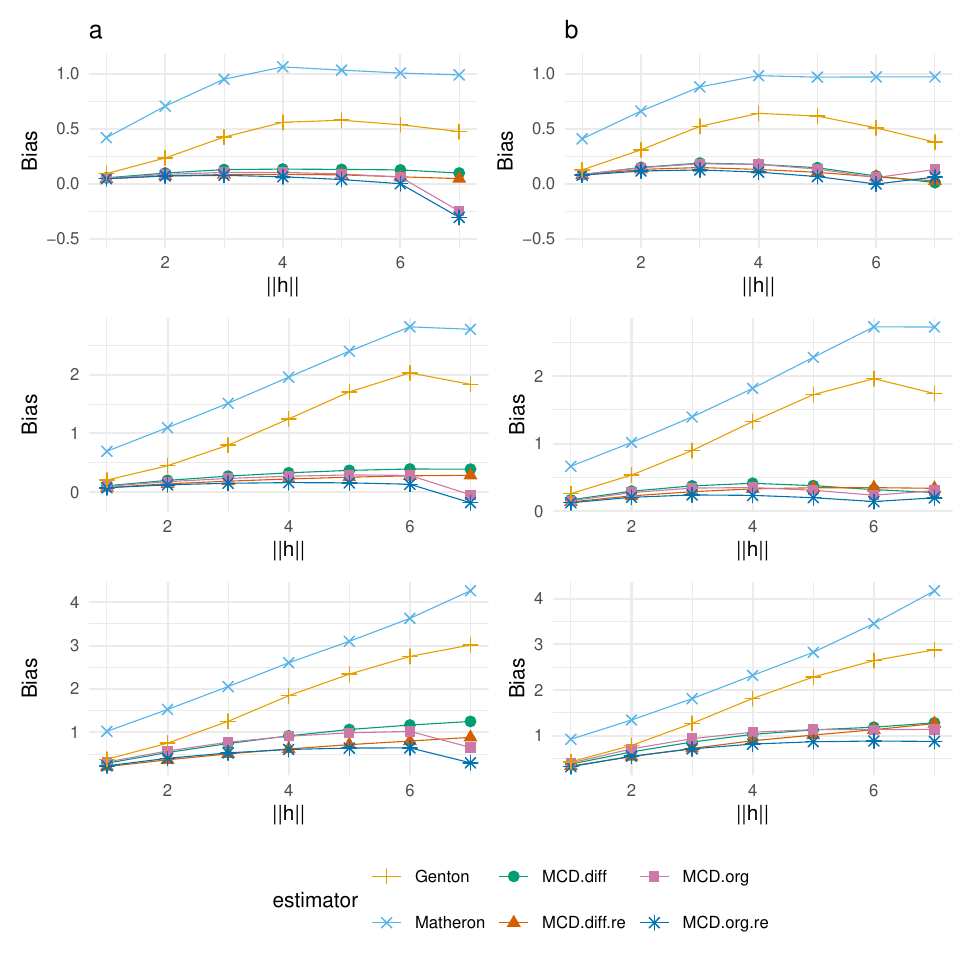}
	\caption{Bias estimation for a $(15\times 15)$ regular grid and an outlier block from the $\mathcal{N}(3, 1)$ distribution with $\epsilon = 5\%$ (first row), $\epsilon = 15\%$ (second row) and $\epsilon = 25\%$ (third row) and for E-W direction (a) and S-N direction (b)}
	\label{fig:block_1}
\end{figure}

Figure \ref{fig:block_1} displays the bias of the different variogram estimators for a $\mathcal{N}(3,1)$ distributed outlier block and a contamination rate of $5\%$ (first row), $15\%$ (second row) and $25\%$ (third row) for the estimation in E-W direction (a) and S-N direction (b). Again we observe that the higher the contamination rate, the larger is the bias of the estimators. The Matheron estimator and the Genton estimator are strongly influenced already by a small outlier block. The differences between the MCD based estimators are not very large, with reweighting reducing their bias. The reweighted MCD.org shows in general the smallest bias. As already seen in Sections \ref{ch:cons} and \ref{ch:oA}, the raw and the reweighted MCD.org have problems estimating the largest lag especially if $||\boldsymbol{h}_{h_{\max}}||$ is smaller than the range of the variogram (E-W direction).

\begin{figure}[H]
	\includegraphics[scale=0.8]{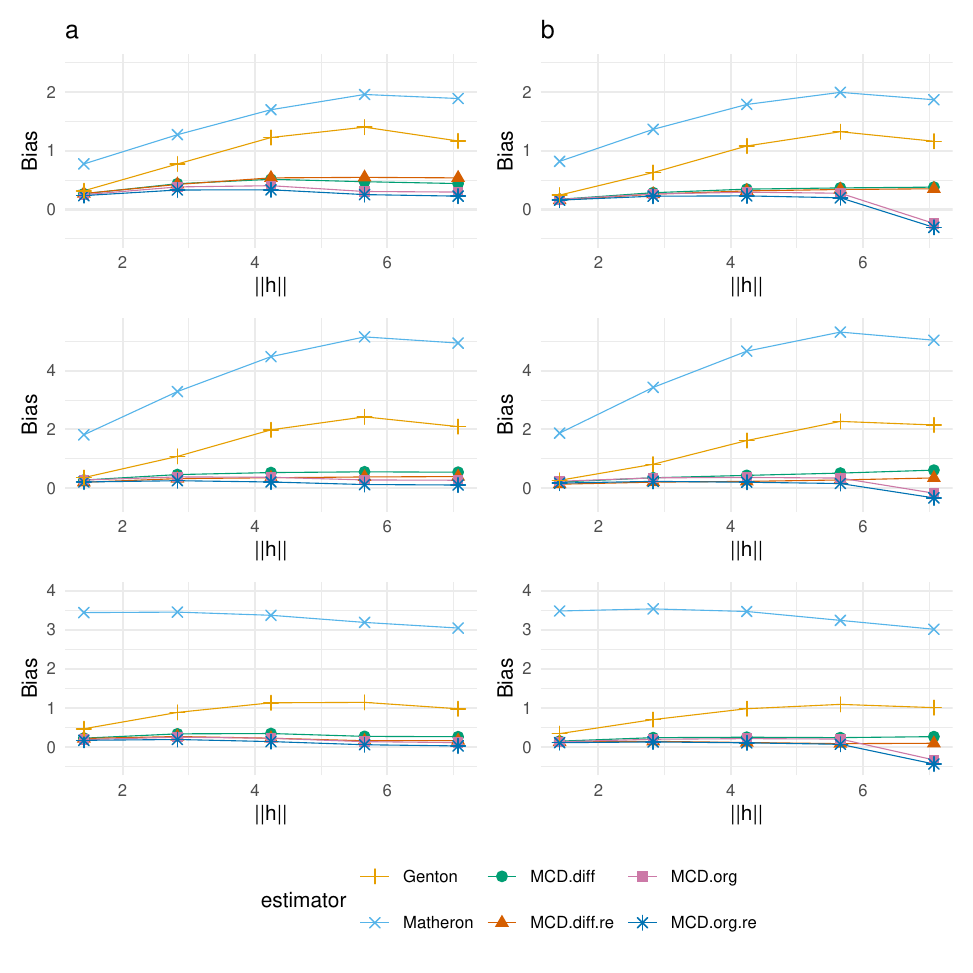}
		\caption{Bias of the  variogram estimation in SW-NE (a) and SE-NW (b) direction per lag of the different estimators for contaminated normally distributed data with a contamination rate of 0.1 and three different contamination distributions: $\mathcal{N}(3, 1)$ (first row), $\mathcal{N}(5,1)$ (second row) and $ \mathcal{N}(0,4)$ (third row)}
		\label{fig:block_2}
\end{figure}

Figure \ref{fig:block_2} depicts the bias of the variogram estimations for estimation in SW-NE(a) and SE-NW (b) direction in case of $10\%$ contamination from three different contamination distributions, namely
$\mathcal{N}(3,1)$  in the first row, $\mathcal{N}(5,1)$ in the second row and $\mathcal{N}(0,4)$ in the third row. Thereby, Figure \ref{fig:block_2} shows the same behavior as Figure \ref{fig:block_1}. In all situations the MCD based estimators are less biased than Matheron and Genton, with reweighting reducing the bias. The bias of the reweighted MCD.org is somewhat smaller than that of the reweighted MCD.diff. As expected, the Matheron estimator always shows the highest bias. Moreover, the bias is for all estimators somewhat smaller for the estimation in SE-NW direction than in SW-NE direction.

%%%%%%%%%%%%%%%%
\subsection{Contamination by isolated outliers}\label{ch:iA}

In this subsection we investigate the behavior of the estimators in case of outliers spread out randomly or at isolated positions. First we consider explosion breakdown for special, but practically relevant scenarios. As before we  concentrate on the one-dimensional case $n_y = 1$, so that  the number of data points is $n=n_x$. 

For the construction of the modified version of the estimators we again assume an $m-$dependent process and build vectors with a minimal distance of $m+1$ between subsequent vectors. Denoting the maximal lag to be estimated by $h_{\max}$, we can thus build $n^{*} = \lfloor \frac{n_x-h_{\max} – 1}{h_{\max}+1+m} +1 \rfloor$ non-overlapping vectors, and each outlier disturbs at most one of these.  
The modified estimator breaks down if  $\lceil\varepsilon_{\text{MCD}}\cdot n^{*}\rceil = n^*-k+1 = n^*-\lfloor \frac{n^{*}+p+1}{2}\rfloor+1$ vectors contain an outlier. The breakdown point thus is 
\begin{align*}
    \varepsilon_{\text{mod}} = \frac{n^*-\lfloor \frac{n^{*}+p+1}{2}\rfloor+1}{n_x}.
\end{align*}
and depends on $h_{\max}$, $n_x$, $m$ and $p$. Again, the larger $h_{\max}$ and $m$, the smaller is $n^{*}$ and the breakdown point. In case of $n_x = 50, m = 1$ and $h_{\max} = 4$, we get $n^{*} = 8$. This leads to the rather small breakdown point  $\varepsilon_{\text{mod}} = \frac{2}{50} = 0.04$ (MCD.org.mod) and $\varepsilon_{\text{mod}} = \frac{3}{50} = 0.06$ (MCD.diff.mod).
Averaging across the different partitions of the data into non-overlapping vectors does not solve these problems. 

Now we discuss breakdown of the estimators MCD.org and MCD.diff in  case of one-dimensional data. As opposed to their modified versions, MCD.org and MCD.diff use overlapping vectors, so that each outlier disturbs up to  $h_{\max} + 1$ vectors, which is the worst case considered here. The number of vectors which can be built is $n^{*} = n_x - h_{\max} $. The estimators can handle less than $\lceil\varepsilon_{\text{MCD}}\cdot n^{*}\rceil = n^*-\lfloor \frac{n^{*}+p+1}{2}\rfloor+1$ contaminated vectors without breakdown. Therefore breakdown requires at least $\frac{n^*-\lfloor \frac{n^{*}+p+1}{2}\rfloor+1}{h_{\max} +1}$ outliers in the data, leading to the following  breakdown point
\begin{align*}
    \varepsilon_{\text{MCD.var}} \geq \nicefrac{\left(\frac{n^*-\lfloor \frac{n^{*}+p+1}{2}\rfloor+1}{h_{\max} +1}\right)}{n_x}.
\end{align*}
This depends on $h_{\max}$ and $n_x$ and is small if $h_{\max}$ is large compared to $n_x$. Considering the example above ($n_x = 50$, $h_{\max} = 4$), we get $n^{*} = 46$ leading to   $\varepsilon_{\text{MCD.var}} = 0.084$ (MCD.org) and  $\varepsilon_{\text{MCD.var}} = 0.088$ (MCD.diff), which is somewhat larger than the breakdown point of the modified estimators. 

\begin{figure}[H]
	\includegraphics[scale=0.8]{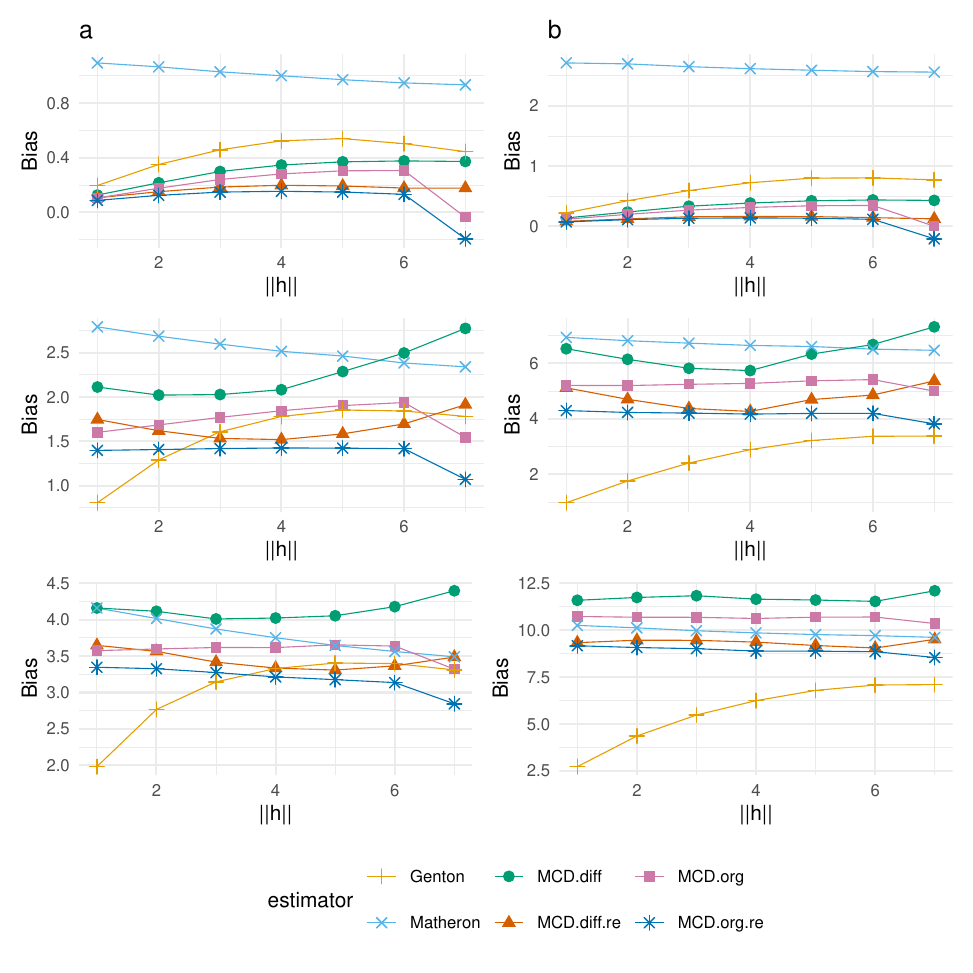}
	\caption{Bias  of the  variogram estimation in E-W direction per lag for several estimators in case of substitutive contamination by a $\mathcal{N}(3,1)$ (a) and $\mathcal{N}(5,1)$ (b) distribution at random positions and contamination rate 0.05 (first row), 0.15 (second row) and 0.25 (third row)}
	\label{fig:iso}
\end{figure}

Breakdown points consider worst case scenarios, and for MCD.org and MCD.mod these are lower bounds. We include some simulations to investigate how the estimators perform in different cases and to compare them with common variogram estimators. In these simulations we consider the following statistical model. The process of interest is the Gaussian Random Field $Z(\boldsymbol{s})$ described at the beginning of Chapter 3. Again we observe a contaminated process $Y(\boldsymbol{s})$ instead of $Z(\boldsymbol{s})$. Here we randomly choose $n_0 = \lceil\epsilon\cdot n \rceil$ different locations from all $n$ locations. Let $N_{n_0}$ be the set with all these locations. Then all locations belonging to $N_{n_0}$ come from the outlier distribution, while all other points come from the process $Z(\boldsymbol{s})$,
\begin{align*}
    Y(\boldsymbol{s})|N_{n_0} = \begin{cases} 
    Z(\boldsymbol{s}),~\text{ if } \boldsymbol{s} \notin N_{n_0} \\
    W,~ \text{ if } \boldsymbol{s} \in N_{n_0}.
    \end{cases}
\end{align*}

Figure \ref{fig:iso} displays the bias of the different variogram estimators for estimation in E-W direction in case of a contaminating $\mathcal{N}(3,1)$ (column a) and $\mathcal{N}(5,1)$ (column b) distribution. In the first row the contamination rate is $5\%$, in the second row it is $15\%$ and in the last row $25\%$. In all situations the Matheron estimator is strongly biased. For $\epsilon = 5\%$  MCD.org, MCD.org.re, MCD.diff and MCD.diff.re are less biased than the Genton estimator. For the larger contamination rates the Genton estimator shows the best results. For a contamination rate of $15\%$ or larger the results for the raw and the reweighted MCD.org and MCD.diff estimators are not comparable to those for the Matheron estimator. The reweighted versions (MCD.org.re and MCD.diff.re) show  a somewhat  smaller bias than the raw ones (MCD.org and MCD.diff). These findings are for moderately large one-sided (positive)  outliers which are located about 2 to 5 standard deviations from the center of the clean data. Further simulations not reported here indicate that the performance of raw and reweighted MCD.org and MCD.diff is much better than this and almost as good as that of the Genton estimator in case of two-sided (both positive and negative) outliers with the same level as the uncontaminated data but a higher variance. 
Summing up, MCD.org and MCD.diff perform good in case of small contamination rates but are worse than the Genton estimator in case of stronger one-sided contamination at random positions.

\section{Application to Satellite Data}\label{ch:app}

Landsat 8 is a satellite belonging to the NASA landsat project. This satellite collects  data of the earth's land surface on 9 different spectral bands in the visible and short-wave infrared spectral regions \citep{Knight2014}. 
We use open source data measured by landsat 8 on October 20, 2016, i.e., a single time point, in a small region of the Brazilian amazon forest. The $x$-coordinates of the region lies between -7332135 and -7319535 and the $y$-coordinates between -1011642 and -995142 in the EPSG:3857 format. This region has a size of 60 x 60 pixels corresponding to 30 times 30 meters each. The data can be downloaded on \url{https://earthexplorer.usgs.gov/}. Our interest here is investigating whether the assumption of isotropy is justified for the Normalized Difference Vegetation Index (NDVI) index.  Therefore we estimate its variogram in all four directions and compare the estimates. The NDVI index is a vegetation index measuring the greenness of biomass. It is calculated using two spectral bands and takes values between -1 and 1. The higher the value of the NDVI, the greener the biomass  \citep{Myneni1995, Tucker1979}. 

\begin{figure}[h]
	\center
	\includegraphics[scale = 0.45]{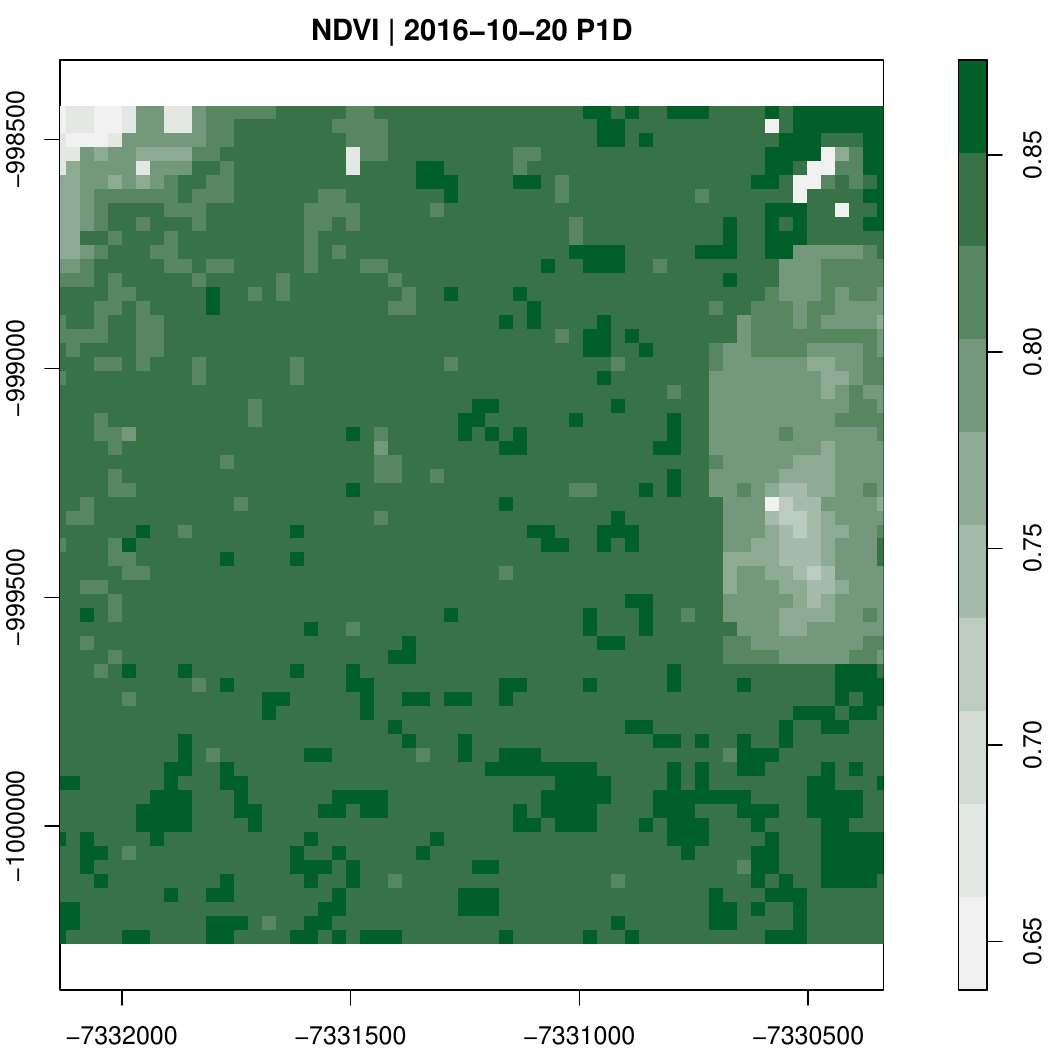}
	\caption{NDVI index in a small region in the Brazilian amazon forest on October 20, 2016 with contamination due to clouds on the right hand side and at the top}
	\label{fig:NDVI}
\end{figure}

The data can be seen in Figure \ref{fig:NDVI}. The NDVI value is quite similar for most pixels, so that the variance of the data is small. Only a block of pixels on the right hand side and another one at the top have much smaller NDVI values compared to the rest. Besides the spectral bands the landsat data also contain a quality band, whichprovides information whether a pixel might be contaminated by clouds, cloud shadows, snow etc., or if it is clear \citep{U.S.GeologicalSurvey2019}. For the pixels showing a different behavior in Figure \ref{fig:NDVI} the quality band indicates clouds or cloud shadows. All other pixels are classified as clear. The data on the right hand side and at the top thus appear to be outlier blocks. 

A simple solution here thus is to use only the data classified as clear for the variogram estimations. In other applications, however, such additional information is not available. We therefore take advantage of the fact that here we have such additional information and compare the variogram estimation based on all data points with the variogram estimation based only on clear data points. We estimate the variogram for the different directions for both data sets using the Matheron estimator, the Genton estimator and both MCD variogram estimators. Due to their better performance in the simulations in Section \ref{ch:sim} we use the reweighted versions of the MCD based estimators. Furthermore in Section \ref{ch:corrfac} we see that for grid sizes larger than $50\times 50$  the correction factors for the MCD based estimators are nearly one and irrespective of the true variogram model. Therefore we do not use these correction factors here. We use $h_{\max} = 4$ for the estimation in E-W and S-N direction and $h_{\max} = 3$ for estimation in the other two directions. Using the clear data without outliers  the grid contains missing values. For the calculation of the MCD based estimators all vectors containing missing values are deleted. Furthermore the variation in the data is very small (the mad of the data with outliers is about $0.009$). To get the variogram estimation on a better scale we standardize both data sets (with and without outliers) using this standard deviation. 

\begin{figure}[H]
    \centering
    \includegraphics[scale=0.8]{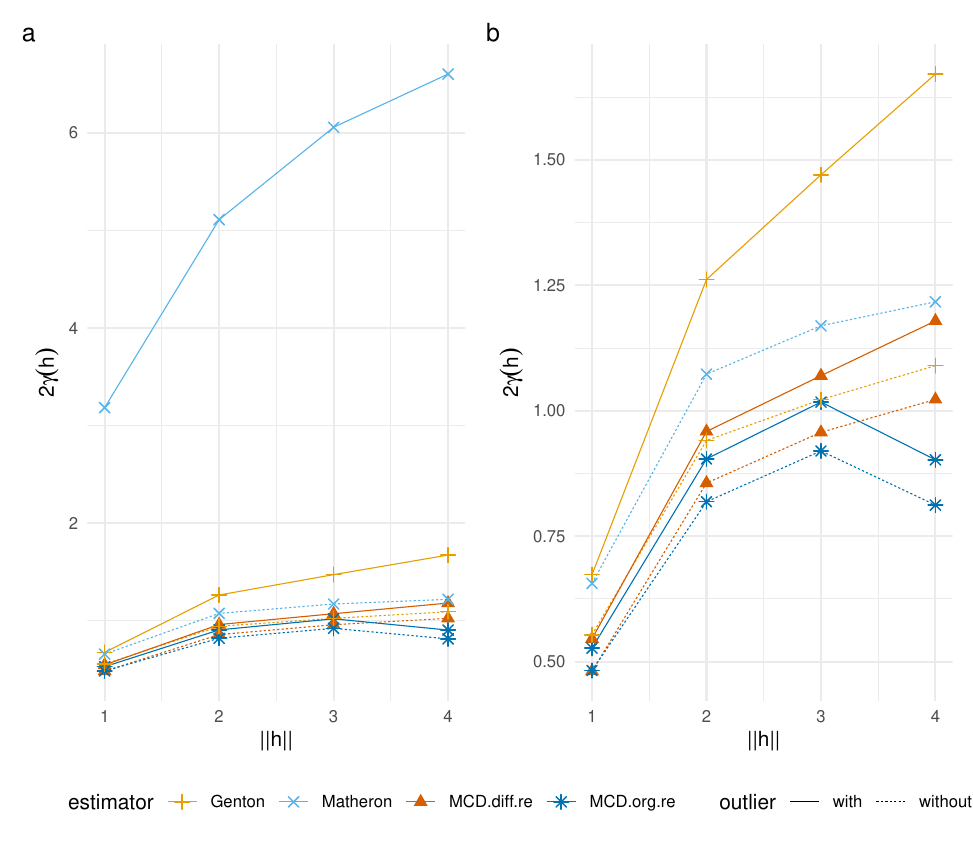}
    \caption{Variogram estimation for the S-N direction in a small region of the Brazilian amazon forest from data with and without outliers. On the left hand side all estimators are shown (a) and on the right hand side the Matheron estimator is excluded for the data with outliers (b)}
    \label{fig:MCD-NDVI}
\end{figure}

Figure \ref{fig:MCD-NDVI} shows the variogram estimations in the S-N direction. On the left hand side (a) we see that the Matheron estimator is strongly influenced by the contaminated pixels. Therefore we exclude this estimator on the right hand side (b) for a better comparison of the other estimators. The Genton estimator is more influenced by the outlier blocks than the MCD based estimators. Using only the clear pixels, both the Genton and the reweighted MCD based estimator provide estimations similar to the Matheron estimation, whereby the Genton estimator provides better results for small lags and MCD.diff.re for large lags. The reweighted MCD based estimator thus can be said to perform well when being applied to the full data set with the outlier block and can therefore be used if no additional information about the quaility of the pixels is available.  

\begin{figure}[H]
    \centering
    \includegraphics[scale=0.8]{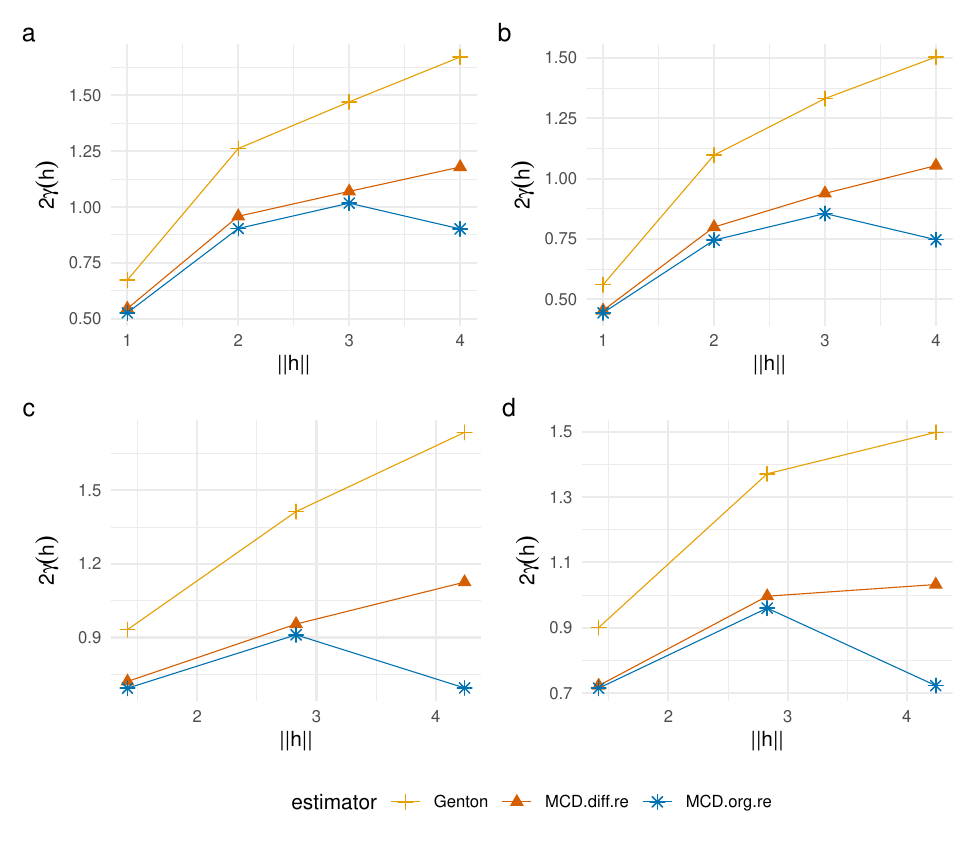}
    \caption{Variogram estimation by the robust estimators in all four direction (S-N: a, E-W: b, SW-NE: c, SE-NW:d) for a small region of the Brazilian amazon forest for data with block outliers}
    \label{fig:comp-NDVI}
\end{figure}

Figure \ref{fig:comp-NDVI} shows the variogram estimation of the robust estimators (Genton, MCD.diff.re, MCD.org.re) for the data with block outliers for all four directions (S-N: a, E-W: b, SW-NE: c, SE-NW:d). The variogram estimations for the different directions are very similar. Accordingly, there seems to be little evidence against the assumption of isotropy here. 

\section{Summary}
The variogram is an important measure of spatial dependency. We have proposed two new estimators which are based on the highly robust multivariate minimum covariance determinant scale estimator (MCD) and compare them to the commonly used variogram estimators by \citet{Matheron1962} and \citet{Genton1998a} in different respects. The new estimators estimate the variogram for several lags jointly. One of them  (MCD.org) assumes data from a weakly stationary random field and applies the MCD to  vectors of subsequent observations to estimate the variance and covariances. 
 The other estimator (MCD.diff) is based on vectors of pairwise differences such that the variogram can be found at the diagonal of the variance-covariance matrix of these vectors. Both estimators can be used for directional variogram estimation in the four main directions (east-west, north-south, southeast-northwest, southwest-northeast). In case of isotropy a joint estimation can be obtained by averaging the estimations for different directions.
 
We see that the Matheron estimator, as expected, performs best in the case of Gaussian Data without outliers. However, the Matheron estimator is strongly influenced by outliers in the data \citep{Lark2000, Kerry2007a} and can thus be recommended only for clean data without outliers.\\
If the data contain isolated outliers, the Genton estimator delivers the best results. However, spatially aggregated outliers occurring in blocks have much stronger effects on the Genton estimator.

The proposed MCD variogram estimators outperform the other estimators in case of outliers occurring in a block. Thereby, reweighting reduces not only the variance of the estimators but also the bias, and MCD.org performs better than MCD.diff regarding bias and rMSE. In case of small grids, unbiased estimation by the reweighted estimators requires correction factors that depend on the unkown true variogram. For the raw estimators the variability of these corrections  is small. A possible approach is using an averaged correction factor based on several plausible parametric models. For increasing grid sizes the variances of MCD.org and MCD.diff decrease and MCD.diff gets unbiased, indicating its asymptotic consistency. MCD.org shows a small bias even for large grid sizes for the largest lag if this lag is smaller than the range of the true variogram. This bias can be reduced by omitting the largest lag from the considerations, but using vectors of unnecessarily high dimension reduces the number of vectors available for estimation and thus increases the variance.  Advantages of the (reweighted) MCD.diff are that it gets unbiased for large grid sizes and requires only the weaker assumption of intrinsic stationarity. 

We recommend the MCD variogram estimators in the possible presence of outlier blocks. Especially for high grid sizes  we recommend usage of the reweighted version. In case of small grid sizes the raw versions may be preferred, since their finite sample correction factors depend much less on the true unkown variogram model. \\
We have concentrated on the case of either one block or outliers at random positions. Further simulations not reported here indicate that the performances of the estimators in mixed situations with both outlier blocks and  isolated outliers are in between.

\backmatter

\bmhead{Acknowledgements}

The authors gratefully acknowledge the computing time provided on the Linux HPC cluster at TU Dortmund University (LiDO3), partially funded in the course of the Large-Scale Equipment Initiative by the German Research Foundation (DFG) as project 271512359. Financial support by the Deutsche Forschungsgemeinschaft (DFG, German Research Foundation; Project-ID 520388526; TRR 391: Spatio-temporal Statistics for the Transition of Energy and Transport) is gratefully acknowledged.

\bmhead{Supplementary information}

An \texttt{R}-implementation for the simulation study in section 4 as well as the \texttt{R}-Code for section 5 is available on GitHub, see \url{https://github.com/JGierse/MCD-variogram-estimator}.

\bmhead{Data availability statements}
The data are open source data and can be downloaded on \url{https://earthexplorer.usgs.gov/}.  The $x$-coordinates of the region lies between -7332135 and -7319535 and the $y$-coordinates between -1011642 and -995142 in the EPSG:3857 format.

\newpage
\begin{appendices}

\section*{Appendix}

\begin{table}[ht]
	\centering
	\begin{tabular}{l|cc}
		grid size & & \\
		$(n_x\times n_y)$ \hspace*{0.5cm}  $n$ & direction & number of vectors \\ \hline 
		\multirow{2}{*}{$(15\times 15)$ \hspace*{0.5cm}  225 }  & E-W/S-N & 120 \\ 
		& SE-NW/SW-NE & 100 \\ \hline
		\multirow{2}{*}{$(25\times 25)$ \hspace*{0.5cm}  625 }  & E-W/S-N & 450 \\ 
		& SE-NW/SW-NE & 400 \\\hline
		\multirow{2}{*}{$(50\times 50)$ \hspace*{0.5cm}  2500 }  & E-W/S-N & 2150 \\ 
		& SE-NW/SW-NE & 2025 \\\hline
		\multirow{2}{*}{$(75\times 75)$ \hspace*{0.5cm}  5625 }  & E-W/S-N & 5100 \\ 
		& SE-NW/SW-NE & 4900\\ \hline
	\end{tabular}
	\caption{Number of vectors available for the multivariate estimations for the different directions and  grid sizes, setting  $h_{\max}=7$ for E-W and S-N, and $h_{\max}=5$ for SE-NW and SW-NE}
	\label{tab:amount_vec}
\end{table}

\begin{table}[ht]
		\centering
		\begin{tabular}{l|rrrr|rrrr}
            & \multicolumn{4}{|c}{exponential} & \multicolumn{4}{|c}{gaussian} \\
			& \multicolumn{4}{|c}{direction} & \multicolumn{4}{|c}{direction} \\ 
			estimator & S-N & E-W & SW-NE & SE-NW & S-N & E-W & SW-NE & SE-NW \\ \hline
            Matheron & 0.99 & 0.99 & 0.99 & 0.99 & 1.01 & 1.01 & 1.02 & 1.00  \\ 
            Genton  & 1.04 & 1.03 & 1.03 & 1.03  & 1.11 & 1.12 & 1.11 & 1.12 \\ 
            MCD.diff  &  0.99 & 0.97 & 0.93 & 0.94  & 1.01 & 1.00 & 0.93 & 0.95 \\ 
            MCD.diff.re  & 1.20 & 1.17 & 1.21 & 1.18 & 1.46 & 1.45 & 1.44 & 1.49 \\ 
            MCD.org  & 0.97 & 0.96 & 0.92 & 0.94 & 0.99 & 1.00 & 0.94 & 0.95 \\ 
            MCD.org.re  & 1.21 & 1.17 & 1.21 & 1.20  & 1.45 & 1.44 & 1.43 & 1.46 \\  
            \hline
		\end{tabular}
	\caption{Simulated finite sample correction factors for the variogram estimators in all four directions in case of a regular 15$\times$ 15 grid for an exponential and an gaussian variogram model}
	\label{tab:corrfac_2}
\end{table}

\newgeometry{bottom = 4cm}
\begin{landscape}
	\begin{table}
		\small
		\begin{tabular}{ccrrrrrrrr|rrrrrrrr}
            & & \multicolumn{8}{c}{E-W} & \multicolumn{8}{|c}{S-N} \\
            %%%%%%%%%%%%%%%%%%%%%%%%%%%%%%%%%%%%%%%%%%%%%%%%%%%%%%%%%%%
			& $\epsilon$ & \multicolumn{4}{c}{0.05} &  \multicolumn{4}{c}{0.15} & \multicolumn{4}{|c}{0.05} & \multicolumn{4}{c}{0.15} \\
            %%%%%%%%%%%%%%%%%%%%%%%%%%%%%%%%%%%%%%%%%%%%%%%%%%%%%%%%%%%
			& distribution & \multicolumn{2}{c}{ $\mathcal{N}(3,1)$} & \multicolumn{2}{c}{ $\mathcal{N}(0,4)$}  & \multicolumn{2}{c}{ $\mathcal{N}(3,1)$}  & \multicolumn{2}{c}{ $\mathcal{N}(0,4)$}  & \multicolumn{2}{|c}{ $\mathcal{N}(3,1)$} & \multicolumn{2}{c}{ $\mathcal{N}(0,4)$}  & \multicolumn{2}{c}{ $\mathcal{N}(3,1)$} & \multicolumn{2}{c}{ $\mathcal{N}(0,4)$} \\ \hline
            %%%%%%%%%%%%%%%%%%%%%%%%%%%%%%%%%%%%%%%%%%%%%%%%%%%%%%%%%%%%
			estimator & lag & $|\text{Bias}|$ & rMSE & $|\text{Bias}|$ & rMSE & $|\text{Bias}|$ & rMSE & $|\text{Bias}|$ & rMSE & $|\text{Bias}|$ & rMSE & $|\text{Bias}|$ & rMSE & $|\text{Bias}|$ & rMSE & $|\text{Bias}|$ & rMSE \\
            %%%%%%%%%%%%%%%%%%%%%%%%%%%%%%%%%%%%%%%%%%%%%%%%%%%%%%%%%%%%
			\multirow{4}{*}{Matheron} & 1 & 4.19 & 4.63 &  17.24 & 19.22 & 6.98 & 7.48 & 49.41 & 51.62 & 4.10 & 4.62 & 17.30 & 19.34 & 6.65 & 7.20 & 49.19 & 51.42  \\
            %%%%%%%%%%%%%%%%%%%%%%%%%%%%%%%%%%%%%%%%%%%%%%%%%%%%%%%%%%%%
            & 4 & 10.64 & 12.57 & 16.34 & 19.07 & 19.57 & 21.83 & 48.83 & 51.75 & 9.84 & 12.89 & 15.93 & 19.31 & 18.17 &  21.07 & 47.99 & 51.23 \\
            %%%%%%%%%%%%%%%%%%%%%%%%%%%%%%%%%%%%%%%%%%%%%%%%%%%%%%%%%%%%
            & 7 & 9.91 & 13.14 & 15.53 & 18.69 & 27.70 & 30.45 & 44.87 & 47.09 & 9.73 & 13.72 & 16.09  & 19.48 & 27.26 & 30.52 & 44.89 & 47.32 \\ \hdashline
            %%%%%%%%%%%%%%%%%%%%%%%%%%%%%%%%%%%%%%%%%%%%%%%%%%%%%%%%%%%%%
			\multirow{4}{*}{Genton} & 1 & 0.93 & 1.12 & 1.14 & 1.33 & 2.09 & 2.25 & 3.61 & 3.77 & 1.27 & 1.72 & 1.71 & 2.11 & 2.59 & 2.96 & 5.30 & 5.69 \\
            %%%%%%%%%%%%%%%%%%%%%%%%%%%%%%%%%%%%%%%%%%%%%%%%%%%%%%%%%%%%
            & 4 & 5.60 & 7.96 & 4.27 & 7.07 & 12.47 & 15.26 & 13.87 & 16.39 & 6.41 & 11.16 & 5.09 & 9.89 & 13.26 & 17.33 & 17.07 &  21.17 \\ 
            %%%%%%%%%%%%%%%%%%%%%%%%%%%%%%%%%%%%%%%%%%%%%%%%%%%%%%%%%%%%
			  & 7 & 4.76 & 9.76 & 3.28 & 8.98 & 18.35 & 22.06 & 16.89 & 20.44 & 3.79 & 9.99 & 2.71 & 9.10 & 17.41 & 21.94 & 16.77 & 21.06 \\ \hdashline
            %%%%%%%%%%%%%%%%%%%%%%%%%%%%%%%%%%%%%%%%%%%%%%%%%%%%%%%%%%%%%
			\multirow{4}{*}{MCD.org} & 1 & 0.51 & 0.97 & 0.49 & 0.97 & 1.01 & 1.37 & 0.92 & 1.28 & 0.87 & 1.73 & 0.91 & 1.72 & 1.54 & 2.22 & 1.62 & 2.29   \\
            %%%%%%%%%%%%%%%%%%%%%%%%%%%%%%%%%%%%%%%%%%%%%%%%%%%%%%%%%%%
            & 4 & 1.02 & 5.62 & 0.70 & 5.82 & 2.69 & 7.14 & 2.10 & 6.57 & 1.77 & 10.70 & 1.20 & 9.76 & 3.57 & 10.74  & 3.50 & 10.96  \\
            %%%%%%%%%%%%%%%%%%%%%%%%%%%%%%%%%%%%%%%%%%%%%%%%%%%%%%%%%%%%%
			& 7 & 2.54 & 7.82 & 2.90 & 8.29 & \underline{0.51} & 8.84 & 1.37 & 8.37  & 1.29 & 12.03 &  0.62 & 11.13 & 3.08 & 12.28 & 2.75 & 12.30\\ \hdashline
            %%%%%%%%%%%%%%%%%%%%%%%%%%%%%%%%%%%%%%%%%%%%%%%%%%%%%%%%%%%%%
			\multirow{4}{*}{MCD.org.mod} & 1 & 1.47 & 1.85 & 3.23 & 4.08 & 2.67 & 2.97 & 5.84 & 6.29 & 1.89 & 2.65 & 5.80 & 7.66 & 3.65 & 4.28 & 10.94 & 12.08 \\
            %%%%%%%%%%%%%%%%%%%%%%%%%%%%%%%%%%%%%%%%%%%%%%%%%%%%%%%%%%%%%
            & 4 & 1.94 & 5.65 & 8.18 & 12.15 & 3.77 & 6.58 & 19.33 & 22.40 & 2.92 & \underline{8.85} & 14.70 & 22.27 & 4.72 & 9.83 & 35.20 & 41.37 \\ 
            %%%%%%%%%%%%%%%%%%%%%%%%%%%%%%%%%%%%%%%%%%%%%%%%%%%%%%%%%%%%%%
			& 7 & -1.97 & 7.10 & 4.31 & 10.79 & \underline{0.30} & \underline{7.05} & 20.56 & 25.68 & 2.23 & 9.96 & 14.02 & 22.49  & 4.87 & 11.21 & 42.60 & 51.11 \\ \hdashline
            %%%%%%%%%%%%%%%%%%%%%%%%%%%%%%%%%%%%%%%%%%%%%%%%%%%%%%%%%%%%%%
	        \multirow{4}{*}{MCD.org.re} & 1 & \underline{0.46} & \underline{0.85} & \underline{0.44} & \underline{0.83} & \underline{0.78} & \underline{1.15} & \underline{0.65} & \underline{1.01} & \underline{0.81} & \underline{1.56} & \underline{0.80} & \underline{1.49} & \underline{1.24} & \underline{1.95} & \underline{1.22} & \underline{1.88}  \\
            %%%%%%%%%%%%%%%%%%%%%%%%%%%%%%%%%%%%%%%%%%%%%%%%%%%%%%%%%%%%%
            & 4 & \underline{0.67} & \underline{4.86} & \underline{0.29} & \underline{4.87} & \underline{1.66} & \underline{6.07} & \underline{0.91} & \underline{5.30} & \underline{1.12} & 9.09 & \underline{0.49} & \underline{8.18} & \underline{2.33} & \underline{9.37} & \underline{1.73} & \underline{8.92}\\
            %%%%%%%%%%%%%%%%%%%%%%%%%%%%%%%%%%%%%%%%%%%%%%%%%%%%%%%%%%%%%%
			& 7 & 3.00 & \underline{7.04} & 3.43 & \underline{7.39} & 1.75 & 7.95 & 2.81 & \underline{7.39} &  0.66 & 10.25 & \underline{0.10} &  9.24 & \underline{1.94} & 10.89 & 0.98 & 10.12 \\ \hdashline
            %%%%%%%%%%%%%%%%%%%%%%%%%%%%%%%%%%%%%%%%%%%%%%%%%%%%%%%%%%%
            \multirow{4}{*}{MCD.diff} & 1 & 0.54 & 1.26 & 0.54 & 1.22 & 1.12 & 1.63 & 1.09 & 1.58 & 0.86 & 2.26 & 0.99 & 2.31 & 1.68 & 2.75 & 1.82 &  2.94 \\
            %%%%%%%%%%%%%%%%%%%%%%%%%%%%%%%%%%%%%%%%%%%%%%%%%%%%%%%%%%%%%
            & 4 & 1.27 & 6.63 & 1.02 & 6.85 & 3.35 & 8.29 & 2.71 & 7.63 & 1.88 & 11.30  & 1.53 & 11.20 & 4.14 & 11.72 & 4.11 &  12.47 \\
            %%%%%%%%%%%%%%%%%%%%%%%%%%%%%%%%%%%%%%%%%%%%%%%%%%%%%%%%%%%%%
			& 7 & 0.88 & 10.01 & \underline{0.48} & 10.69 & 3.86 & 12.54 & 2.29 &  10.75 & \underline{0.20} & 11.20 & 0.69 & 10.59 & 2.84 & 13.23 & 2.13 & 12.41 \\ \hdashline
            %%%%%%%%%%%%%%%%%%%%%%%%%%%%%%%%%%%%%%%%%%%%%%%%%%%%%%%%%%%%%
			\multirow{4}{*}{MCD.diff.mod} & 1 & 1.31 & 1.98 & 2.24 & 3.54 & 2.22 & 3.03 & 3.48 & 4.71 &  1.65 & 2.94 & 4.44 & 7.71 &  3.35 & 4.94 & 7.26 & 11.23 \\
            %%%%%%%%%%%%%%%%%%%%%%%%%%%%%%%%%%%%%%%%%%%%%%%%%%%%%%%%%%%%
            & 4 & 1.81 & 6.11 & 6.76 & 10.98 & 3.15 & 6.63 & 16.41 & 20.39 & 2.67 & 9.61 & 14.50 & 23.53 & 4.22 & 10.24 &  34.52 & 43.27  \\
            %%%%%%%%%%%%%%%%%%%%%%%%%%%%%%%%%%%%%%%%%%%%%%%%%%%%%%%%%%%%
			& 7 & 0.87 & 9.00 & 6.50 & 12.50 & 2.63 & 9.08 & 22.64 & 26.98 & 0.47 & 9.55 & 13.47 & 22.20 & 3.77 & \underline{10.40} & 47.77 & 55.87 \\ \hdashline
            %%%%%%%%%%%%%%%%%%%%%%%%%%%%%%%%%%%%%%%%%%%%%%%%%%%%%%%%%%%
			\multirow{4}{*}{MCD.diff.re} & 1 & \underline{0.46} & 0.97 & 0.47 & 0.96 & 0.82 & 1.32 & 0.76 & 1.22 & 0.84 & 1.85 & 0.89 & 1.84 & 1.36 & 2.38 & 1.41 & 2.41\\
            %%%%%%%%%%%%%%%%%%%%%%%%%%%%%%%%%%%%%%%%%%%%%%%%%%%%%%%%%%%
            & 4 & 0.78 & 5.24  &0.43 & 5.23 & 2.27 & 6.91  & 1.17 &  5.76 &1.43 & 9.49 & 0.82 & 9.00 & 3.39 & 10.44 & 2.13 & 9.81 \\
            %%%%%%%%%%%%%%%%%%%%%%%%%%%%%%%%%%%%%%%%%%%%%%%%%%%%%%%%%%%%%
            & 7 & \underline{0.40} & 8.18 & 0.40 & 8.25 & 2.88 & 10.90 & \underline{0.31} & 8.40 & 0.22 & \underline{9.58} & 0.95 & \underline{8.61} & 3.42 & 12.42 & \underline{0.62} & \underline{10.08} \\
			
		\end{tabular}
	\caption{Bias and rMSE for the estimation in E-W and S-N direction in case of different contamination rates $\epsilon$ and different contamination distributions for three different lags. All values are multiplied by 10 for ease of presentation. The best value is underlined in each case}
	\label{tab:mA}
	\end{table}
\end{landscape}	
 \restoregeometry

\end{appendices}

\bibliography{library}% common bib file

\begin{thebibliography}{33}
\providecommand{\natexlab}[1]{#1}
\providecommand{\url}[1]{{#1}}
\providecommand{\urlprefix}{URL }
\providecommand{\doi}[1]{\url{https://doi.org/#1}}
\providecommand{\eprint}[2][]{\url{#2}}
 \bibcommenthead

\bibitem[{Bowman and Crujeiras(2013)}]{Bowman2013}
Bowman AW, Crujeiras RM (2013) Inference for variograms. Computational
  Statistics and Data Analysis 66:19--31. \doi{10.1016/j.csda.2013.02.027}

\bibitem[{Butler et~al(1993)Butler, Davies, and Jhun}]{Butler1993}
Butler RW, Davies PL, Jhun M (1993) Asymptotics for the minimum covaraince
  determinant estimator. The Annals of Statistics 21:1385--1400.
  \doi{10.1214/aos/1176349264}

\bibitem[{Cressie and Hawkins(1980)}]{Cressie1980}
Cressie N, Hawkins DM (1980) Robust estimation of the variogram: I. Journal of
  the International Association for Mathematical Geology 12:115--125.
  \doi{10.1007/BF01035243}

\bibitem[{Croux and Haesbroeck(1999)}]{Croux1999}
Croux C, Haesbroeck G (1999) Influence function and efficiency of the minimum
  covariance determinant scatter matrix estimator. Journal of Multivariate
  Analysis 71:161--190. \doi{10.1006/jmva.1999.1839}

\bibitem[{Dürre et~al(2015)Dürre, Fried, and Liboschik}]{Durre2015}
Dürre A, Fried R, Liboschik T (2015) Robust estimation of (partial)
  autocorrelation. WIREs Computational Statistics 7:205--222.
  \doi{10.1002/wics.1351}

\bibitem[{Garthoff and Otto(2017)}]{Garthoff2017}
Garthoff R, Otto P (2017) Control charts for multivariate spatial
  autoregressive models. AStA Advances in Statistical Analysis 101:67--94.
  \doi{10.1007/s10182-016-0276-x}

\bibitem[{Genton(1998{\natexlab{a}})}]{Genton1998a}
Genton MG (1998{\natexlab{a}}) Highly robust variogram estimation. Mathematical
  Geology 30:213--221. \doi{10.1023/A:1021728614555}

\bibitem[{Genton(1998{\natexlab{b}})}]{Genton1998}
Genton MG (1998{\natexlab{b}}) Spatial breakdown point of variogram estimators.
  Mathematical Geology 30:853--871. \doi{10.1023/A:1021778626251}

\bibitem[{Guan et~al(2004)Guan, Sherman, and Calvin}]{Guan2004}
Guan Y, Sherman M, Calvin JA (2004) A nonparametric test for spatial isotropy
  using subsampling. Journal of the American Statistical Association
  99:810--821. \doi{10.1198/016214504000001150}

\bibitem[{Hawkins and Olive(2002)}]{Hawkins2002}
Hawkins DM, Olive DJ (2002) Inconsistency of resampling algorithms for
  high-breakdown regression estimators and a new algorithm. Journal of the
  American Statistical Association 97:136--159.
  \doi{10.1198/016214502753479293}

\bibitem[{Hubert and Debruyne(2010)}]{Hubert2010}
Hubert M, Debruyne M (2010) Minimum covariance determinant. Wiley
  Interdisciplinary Reviews: Computational Statistics 2:36--43.
  \doi{10.1002/wics.61}

\bibitem[{Hubert et~al(2012)Hubert, Rousseeuw, and Verdonck}]{Hubert2012}
Hubert M, Rousseeuw PJ, Verdonck T (2012) A deterministic algorithm for robust
  location and scatter. Journal of Computational and Graphical Statistics
  21:618--637. \doi{10.1080/10618600.2012.672100}

\bibitem[{Kerry and Oliver(2007)}]{Kerry2007a}
Kerry R, Oliver MA (2007) Determining the effect of asymmetric data on the
  variogram. ii. outliers. Computers and Geosciences 33:1233--1260.
  \doi{10.1016/j.cageo.2007.05.009}

\bibitem[{Knight and Kvaran(2014)}]{Knight2014}
Knight EJ, Kvaran G (2014) Landsat-8 operational land imager design,
  characterization and performance. Remote Sensing 6:10286--10305.
  \doi{10.3390/rs61110286}

\bibitem[{Lark(2008)}]{Lark2008}
Lark MR (2008) Some results on the spatial breakdown point of robust point
  estimates of the variogram. Mathematical Geosciences 40:729--751.
  \doi{10.1007/s11004-008-9162-8}

\bibitem[{Lark(2000)}]{Lark2000}
Lark RM (2000) A comparison of some robust estimators of the variogram for use
  in soil survey. European Journal of Soil Science 51:137--157.
  \doi{10.1046/j.1365-2389.2000.00280.x}

\bibitem[{Lopuhaä(1999)}]{Lopuhaa1999}
Lopuhaä HP (1999) Asymptotics of reweighted estimators of multivariate
  location and scatter. Annals of Statistics 27:1638--1665.
  \doi{10.1214/aos/1017939145}

\bibitem[{Lopuhaä and Rousseeuw(1991)}]{Lopuhaa1991}
Lopuhaä HP, Rousseeuw PJ (1991) Breakdown points of affine equivariant
  estimators of multivariate location and covariance matrices. The Annals of
  Statistics 19:229--248. \doi{doi:10.1214/aos/1176347978}

\bibitem[{Maechler et~al(2021)Maechler, Rousseeuw, Croux, Todorov, Ruckstuhl,
  Salibian-Barrera, Verbeke, Koller, Conceicao, and di~Palma}]{robustbase2021}
Maechler M, Rousseeuw PJ, Croux C, et~al (2021) robustbase: Basic robust
  statistics. \urlprefix\url{http://robustbase.r-forge.r-project.org/}

\bibitem[{Matheron(1962)}]{Matheron1962}
Matheron G (1962) Traité de géostatistique appliquée, Tome I. Mémoires du
  Bureau de Recherches Géologiques et Minières

\bibitem[{Myneni et~al(1995)Myneni, Hall, Sellers, and Marshak}]{Myneni1995}
Myneni RB, Hall FG, Sellers PJ, et~al (1995) The interpretation of spectral
  vegetation indexes. IEEE Transactions on Geoscience and Remote Sensing
  33:481--486. \doi{10.1109/36.377948}

\bibitem[{Otto and Schmid(2016)}]{Otto2016}
Otto P, Schmid W (2016) Detection of spatial change points in the mean and
  covariances of multivariate simultaneous autoregressive models. Biometrical
  Journal 58:1113--1137. \doi{10.1002/bimj.201500148}

\bibitem[{Pison et~al(2002)Pison, Aelst, and Willems}]{Pison2002}
Pison G, Aelst SV, Willems G (2002) Small sample corrections for lts and mcd.
  Metrika 55:111--123. \doi{10.1007/s001840200191}

\bibitem[{{R Core Team}(2024)}]{r2024}
{R Core Team} (2024) R: A Language and Environment for Statistical Computing. R
  Foundation for Statistical Computing, Vienna, Austria,
  \urlprefix\url{https://www.R-project.org/}

\bibitem[{Roelant et~al(2009)Roelant, Aelst, and Willems}]{Roelant2009}
Roelant E, Aelst SV, Willems G (2009) The minimum weighted covariance
  determinant estimator. Metrika 70:177--204. \doi{10.1007/s00184-008-0186-3}

\bibitem[{Rousseeuw(1985)}]{Rousseeuw1985}
Rousseeuw PJ (1985) Multivariate estimation with high breakdown point.
  Mathematical Statistics and Applications Vol B pp 283--297.
  \doi{10.1007/978-94-009-5438-0_20}

\bibitem[{Rousseeuw and Croux(1993)}]{Rousseeuw1993}
Rousseeuw PJ, Croux C (1993) Alternatives to the median absolute deviation.
  Journal of the American Statistical Association 88:1273--1283.
  \doi{10.1080/01621459.1993.10476408}

\bibitem[{Rousseeuw and Driessen(1999)}]{Rousseeuw1999}
Rousseeuw PJ, Driessen KV (1999) A fast algorithm for the minimum covariance
  determinant estimator. Technometrics 41:212--223.
  \doi{10.1080/00401706.1999.10485670}

\bibitem[{Schlather et~al(2020)Schlather, Malinowski, Oesting, Boecker,
  Strokorb, Engelke, Martini, Ballani, Moreva, Auel, Menck, Gross, Riberio,
  Ripley, Singelton, Pfaff, and Team}]{randomfiels2020}
Schlather M, Malinowski A, Oesting M, et~al (2020) Randomfields: Simulation and
  analysis of random fields.
  \urlprefix\url{https://cran.r-project.org/package=RandomFields}

\bibitem[{Sherman(2011)}]{Sherman2011}
Sherman M (2011) Spatial Statistics and Spatio-Temporal Data, 1st edn. John
  Wiley and Sons

\bibitem[{Todorov(1992)}]{Todorov1992}
Todorov V (1992) Computing the minimum covariance determinant estimator (mcd)
  by simulated annealing. Computational Statistics and Data Analysis
  14:515--525. \doi{10.1016/0167-9473(92)90067-P}

\bibitem[{Tucker(1979)}]{Tucker1979}
Tucker CJ (1979) Red and photographic infrared linear combinations for
  monitoring vegetation. Remote Sensing of Environment 8:127--150.
  \doi{10.1016/0034-4257(79)90013-0}

\bibitem[{{U.S. Geological Survey}(2019)}]{U.S.GeologicalSurvey2019}
{U.S. Geological Survey} (2019) Landsat 8 data users handbook. Nasa
  \urlprefix\url{https://landsat.usgs.gov/documents/Landsat8DataUsersHandbook.pdf}

\end{thebibliography}

\end{document}